\documentclass[reprint,superscriptaddress,nofootinbib,amsmath,amssymb]{revtex4-1}
\usepackage{xcolor}
\usepackage{cancel}
\usepackage{graphicx}
\usepackage[normalem]{ulem}

\newtheorem{thm}{THEOREM}[section]

\newtheorem{prp}[thm]{PROPOSITION}
\newtheorem{dfn}[thm]{Definition}
\newtheorem{rem}[thm]{Remark}

\newcommand{\eps}{\epsilon}
\newcommand{\veps}{\varepsilon}

\newcommand{\drm}{\mathrm{d}}

\newcommand{\Ddt}{\frac{\drm\phantom{s}}{\drm t}}

\newcommand{\pddt}{\frac{\partial\phantom{t}}{\partial t}}
\newcommand{\tpddt}{{\textstyle{\frac{\partial\phantom{t}}{\partial t}}}}

\newcommand{\pddct}{{{\frac{1}{c}\frac{\partial\phantom{t}}{\partial t}}}}

\newcommand{\pddtSQUARE}{{\textstyle{\frac{\partial^2\phantom{t}}{\partial t^2}}}}

\newcommand{\ONE}{{\boldsymbol{1}}}

\newcommand{\refeq}[1]{(\ref{#1})}

\newcommand{\mbare}{m_{\text{b}}}

\newcommand{\vect}[1] {\boldsymbol{{ #1}} }

\newcommand{\qv}[1]{{\textbf{\textsf{#1}}}}

\newcommand{\tenseur}[1]{{\textbf{\textsf{#1}}}}

\newcommand{\Rset}{\mathbb{R}}

\newcommand{\FQ}{\tenseur{F}}           
\newcommand{\gQ}{\tenseur{g}}           

\newcommand{\fQ}{\qv{f}}        	
\newcommand{\qQ}{\qv{q}}                

\newcommand{\aV}{\vect{a}}              

\newcommand{\fV}{\vect{f}}              
\newcommand{\gV}{\vect{\Phi}}              
\newcommand{\jV}{{\vect{j}}}		
\newcommand{\nV}{{\vect{n}}}		
\newcommand{\pV}{{\vect{p}}}            
\newcommand{\qV}{{\vect{q}}}            
\newcommand{\QV}{{\vect{Q}}}            
\newcommand{\sV}{{\vect{s}}}            
\newcommand{\vV}{{\vect{v}}}            

\newcommand{\BLWr}{{\BV}_{\text{\textsc{lw}}}^{\text{\tiny{ret}}}} 
\newcommand{\DLWr}{{\DV}_{\text{\textsc{lw}}}^{\text{\tiny{ret}}}} 
\newcommand{\ELWr}{{\EV}_{\text{\textsc{lw}}}^{\text{\tiny{ret}}}} 
\newcommand{\HLWr}{{\HV}_{\text{\textsc{lw}}}^{\text{\tiny{ret}}}} 

\newcommand{\NullV}{\vect{0}}
\newcommand{\AV}{\pmb{{\cal A}}}
\newcommand{\BV}{\pmb{{\cal B}}}

\newcommand{\DV}{\pmb{{\cal D}}}
\newcommand{\EV}{\pmb{{\cal E}}}

\newcommand{\HV}{\pmb{{\cal H}}}
\newcommand{\ZV}{\pmb{{\cal Z}}}

\newcommand{\cA}{{\mathcal{A}}}

\newcommand{\cV}{{\mathcal{V}}}

\newcommand{\PiV}{\boldsymbol{\Pi}}
\newcommand{\PV}{\vect{P}}

\newcommand{\nab}{\vect{\nabla}}

\newcommand{\abs}[1]{\big\vert #1 \big\vert}

\renewcommand{\leq}{\leqslant}
\renewcommand{\geq}{\geqslant}

\newcommand{\cL}{{\cal L}} 

\newcommand{\crprd}{{\boldsymbol\times}}

\begin{document}
	
\title{Force on a point charge source of the classical electromagnetic field}
	
\author{Michael K.-H. Kiessling}
\email{miki@math.rutgers.edu}
\affiliation{Department of Mathematics, Rutgers University,
                110 Frelinghuysen Rd., Piscataway, NJ 08854, USA}
	
\begin{abstract}
\noindent 
 It is shown that a well-defined expression for the total electromagnetic force $\fV^{\mbox{\tiny{em}}}$ on 
a point charge source of the classical electromagnetic field can be extracted from the postulate of total 
momentum conservation whenever the classical electromagnetic field theory satisfies a handful of regularity conditions. 
 Among these is the generic local integrability of the field momentum density over a neighborhood of the point charge.
 This disqualifies the textbook Maxwell--Lorentz
field equations, while the Maxwell--Bopp--Land\'e--Thomas--Podolsky 
field equations qualify, and presumably so do the Maxwell--Born--Infeld 
field equations. 
 Most importantly, when the usual relativistic relation between the velocity and the momentum of a point charge with 
``bare rest mass'' $\mbare\neq 0$ is postulated, Newton's law $\Ddt\pV = \fV$ with $\fV=\fV^{\mbox{\tiny{em}}}$
becomes an integral equation for the point particle's acceleration; the infamous third-order time derivative of 
the position which plagues the Abraham--Lorentz--Dirac equation of motion does not show up. 
 No infinite bare mass renormalization is invoked, and no ad hoc averaging of fields over a neighborhood of the
point charge.
 The approach lays the rigorous microscopic foundations of classical electrodynamics with point charges.
\begin{description}
 \item[Preprint with erratum implemented (colored correction markings)] April 21, 2020
\end{description}
\end{abstract}

\maketitle
	
\section{Introduction}
 The practical success of the Lorentz formula \cite{LorentzFORCE} 
\begin{equation}
\fV^{\mbox{\tiny{em}}}_{\mbox{\tiny{Lor}}}(t)
= \label{eq:Lf}
  -e\left[\EV(t,\qV(t)) + \textstyle{\frac{1}{c}}\vV(t) \crprd\BV(t,\qV(t))\right]
\end{equation}
for the electromagnetic force exerted by a \emph{given, smooth} 
electric field  $\EV(t,\sV)$ and magnetic induction field $\BV(t,\sV)$ 
on a moving \emph{test point electron} with charge $-e$, position  $\qV(t)$, and velocity  $\vV(t)$ is well established;
here, $c$ is the speed of light in vacuum.
 However, this formula is notoriously ill-defined when the point electron is not idealized
as a ``test particle'' but treated properly as a source of the electromagnetic fields with which it interacts. 
 The Maxwell--Lorentz  equations for these fields, consisting of the two evolution equations
\begin{alignat}{1}
\textstyle
\pddt{\BV(t,\sV)}
&= \label{eq:MLdotB}
        - c \nab\crprd\EV(t,\sV) \, ,
\\
\textstyle
\pddt{\EV(t,\sV)}
&= 
        + c\nab\crprd\BV(t,\sV)  + 4\pi e {\vV}(t)\delta_{\qV(t)}(\sV)\, ,
\label{eq:MLdotE}
\end{alignat}
together with the two constraint equations 
\begin{alignat}{1}
        \nab\cdot \BV(t,\sV)  
&= \label{eq:MLdivB}
        0\, ,
\\
        \nab\cdot\EV(t,\sV)  
&=
        -4 \pi  e \delta_{\qV(t)}(\sV)\, ,
\label{eq:MLdivE}
\end{alignat}
make it plain that Maxwell--Lorentz (ML) fields with a single point charge source, at $\qV(t)$, \emph{must}
have some singularity at $\sV=\qV(t)$. 
 In the remainder of this introduction we first summarize the current state of affairs in dealing with this problem,
and then we recall the major deficiencies of this approach that were pointed out by others already.
 The rest of this paper is devoted to setting up a well-defined classical theory of point charge motion.
 
 We will occasionally invoke a manifestly co-variant geometrical four-vector notation, but mostly we will work with the 
space \&\ time splitting well-suited for the formulation of a dynamical initial value problem we seek, 
even though the Lorentz co-variance is then obscured.
 Note that the above formulas are valid in any flat foliation of Minkowski spacetime into Euclidean space points 
$\sV\in\Rset^3$ at time $t\in\Rset$.

\vspace{-10pt}
                \subsection{The current state of affairs}\vspace{-5pt}
 The field singularity associated with a motion $t\mapsto\qV(t)$ having bounded piecewise continuous acceleration
$\aV(t)$ has been known explicitly for a long time.
 Since the system of Maxwell--Lorentz field equations is linear, 
their general distributional solution can be written as the sum of the general Lipschitz continuous source-free 
electromagnetic field solution, for which \refeq{eq:Lf} makes perfect sense, 
plus the retarded Li\'enard--Wiechert field \cite{lienard}, \cite{Wiechert}, $\BLWr:=\HLWr$ \&\ $\ELWr:=\DLWr$, with
(cf. \cite{JacksonBOOKb})
\begin{alignat}{1}
\hskip-.6truecm
{\DLWr(t,\sV)} &=\label{eq:LWsolE}
  -e \frac{c^2-|\vV|^2}{|\sV-\qV|^2} 
\frac{{c\nV(\qV,\sV)}_{\phantom{!\!}}-{\vV}}{\bigl(\textstyle{c-\nV(\qV,\sV)\cdot {\vV}}\bigr)^{\!3}} 
\Biggl.\Biggr|_{\mathrm{ret}}\hskip-1truecm
\\
&\quad \notag
-e 
\frac{\nV(\qV,\sV)\crprd
\bigl[\bigl(c\nV(\qV,\sV)_{\phantom{!\!}}-{\vV}\bigr)\crprd\aV\bigr]}{{|\sV-\qV|}\bigl(\textstyle{c-\nV(\qV,\sV)\cdot {\vV}}\bigr)^{\!3}}
\Biggl.\Biggr|_{\mathrm{ret}}\hskip-1truecm
\\
\hskip-1truecm
{\HLWr(t,\sV)} 
&= \label{eq:LWsolB}
        \nV(\qV,\sV)|_{_{\mathrm{ret}}}\crprd {\DLWr(t,\sV)}
\, ,\vspace{-10pt}
\end{alignat}
where $\nV(\qV,\sV) = \frac{\sV-\qV}{|\sV-\qV|}$ is a \emph{normalized} vector from $\qV$ to $\sV$,
and where ``$|_{\mathrm{ret}}$'' means that $(\qV,\vV,\aV)= (\qV,\vV,\aV)(t^{\mathrm{ret}})$ 
with $t^{\mathrm{ret}}(t,\sV)$ being defined implicitly by $c(t-t^{\mathrm{ret}}) = |\sV-\qV(t^{\mathrm{ret}})|$;
here, $\aV$ is the acceleration vector of the point charge.
 The electromagnetic Li\'enard--Wiechert fields $\BLWr$ and $\ELWr$ exhibit both
a $\propto 1/r^2$ and a $\propto 1/r$ singularity, where $r$ denotes $|\sV-\qV(t)|$; they each
have a directional singularity at the location of the  point charge source, too.

 In an attempt to give some mathematical meaning to the manifestly ill-defined symbolic expressions
``$\EV(t,\qV(t))$'' and ``$\BV(t,\qV(t))$'' when $\EV(t,\sV)$ and $\BV(t,\sV)$ are a sum of a regular source-free
field and  the Li\'enard--Wiechert fields \refeq{eq:LWsolE} and \refeq{eq:LWsolB}, Lorentz and his contemporaries
averaged the fields $\EV(t,\sV)$ and $\BV(t,\sV)$ over a neighborhood of the point charge at $\qV(t)$, but this
does not lead to unambiguous finite vector values for ``$\EV(t,\qV(t))$'' and ``$\BV(t,\qV(t))$,'' and when the 
neighborhood is shrunk to the point $\qV(t)$ the infinities are back.
 The conclusion at the time (and also more recently in \cite{GHW}) was that the physical electron cannot be assumed to be a point,
but must have an extended charge distribution and perhaps some other structure, all of which to determine became 
a goal of what Wiechert and Lorentz called ``electron theory'' \cite{LorentzENCYCLOP}, nowadays referred to as ``classical electron theory.'' 
 It's an interesting dynamical theory in its own right; for more recent investigations, see the books \cite{YaghjianBOOK} and 
\cite{Spohn}, and the papers \cite{GHW} and \cite{AppKieAOP}.
 Since we are interested in the theory of point charge motion, we here do not spend much time with classical electron 
theory, except that we note that some insights gained in its pursuit made their way into the prevailing classical theory 
of point electron motion which was put together by Dirac and Landau \&\ Lifshitz.

 In 1938 Dirac \cite{DiracA} invented \emph{negative infinite bare mass renormalization} to avoid the infinities 
which occur when the averaging surface about $\qV(t)$ is shrunk to $\qV(t)$. 
 Following Fermi's contribution to classical electron theory \cite{Fermi}, Dirac averaged the fields over a sphere of 
radius $r$ centered at the electron in its instantaneous rest-frame.
 Also, he worked with linear combinations of the retarded and advanced representations of the fields.
 With $m_{\mbox{\tiny{obs}}}$ denoting the electron's ``observable rest mass,''
Dirac assigned an \emph{averaging radius-dependent} bare mass  ${\mbare}(r)$ to the point electron, defined by
\begin{equation}
{\textstyle m_{\mbox{\tiny{obs}}} = \lim_{r\downarrow 0}\left({\mbare} (r) +\frac{e^2}{2c^2}\frac1r\right)};
\end{equation}
evidently, ${\mbare}(r)\downarrow -\infty$ as $r\downarrow 0$. 
  As is well-known, Dirac's mass-renormalization computations became the template for the
modern renormalization group approach to quantum electrodynamics and, more generally, quantum field theory.
 However, if electrons are true points without structure, then Dirac's construction is 
logically incomprehensible: if a point electron has a bare mass, then it cannot depend on the
radius $r$ of a sphere over which one averages the Maxwell--Lorentz fields.

 Postponing such logical concerns until later in his life, Dirac obtained the Abraham--Lorentz--Dirac equation, which
in the four-vector notation explained in \cite{AppKieAOP} reads
\begin{alignat}{1}\label{AbrLorDiracEQ}
\textstyle
m_{\mbox{\tiny{obs}}}
\frac{\drm^2}{\drm\tau^2}\qQ
= 
& -\frac{e}{c}\FQ^{\mbox{\tiny{ext}}}(\qQ)\cdot\frac{\drm}{\drm\tau}\qQ \\ \notag
&+ \frac{2e^2}{3c^3} 
\left(\gQ +\frac{1}{c^2}\frac{\drm}{\drm\tau}\qQ \otimes\frac{\drm}{\drm\tau}\qQ\right)\cdot
{\textstyle\frac{\drm^3}{\drm\tau^3}\qQ }  
\end{alignat}
where 
the term in the first line at r.h.s.\refeq{AbrLorDiracEQ}
is an ``externally generated'' test-particle-type {Lorentz Minkowski}-force, and
the term in the second line at r.h.s.\refeq{AbrLorDiracEQ}
is {von Laue}'s radiation-reaction {Minkowski}-force.

 While \refeq{AbrLorDiracEQ} is free of infinities if $\FQ^{\mbox{\tiny{ext}}}(\qQ)$ is smooth, 
the third proper time derivative in the {von Laue Minkowski}-force 
means that \refeq{AbrLorDiracEQ} is a third-order ODE for the position of the particle as a function of (proper) time.
 The pertinent initial value problem therefore requires vector initial data for position, velocity, and acceleration. 
 Yet a classical initial value problem of point particle motion \emph{may only involve initial data for position and velocity}.

 {Landau} \&\ {Lifshitz} \cite{LandauLifshitz} handled this $\dddot\qQ$ problem in the
following perturbative manner. 
 They argued that {von Laue}'s  $\dddot\qQ$ {Minkowski-force term must be a small perturbation} of $\fQ^{\mbox{\tiny{ext}}}$
whenever test-particle theory works well. 
 In such situations one may compute $\dddot\qQ$ perturbatively by taking the proper time derivative of the test-particle 
equation of motion, i.e.
\begin{equation}\label{dddq}
\textstyle
\frac{\drm^3}{\drm\tau^3}\qQ  \approx -\frac{e}{m_{\mbox{\tiny{obs}}}c}\frac{\drm}{\drm\tau}
\left(\FQ^{\mbox{\tiny{ext}}}(\qQ)\cdot\frac{\drm}{\drm\tau}\qQ\right).
\end{equation}
 R.h.s.\refeq{dddq} depends only on $\qQ$, $\dot\qQ$, $\ddot\qQ$.
 If we substitute it for $\frac{\drm^3}{\drm\tau^3}\qQ$ at r.h.s.\refeq{AbrLorDiracEQ}, equation \refeq{AbrLorDiracEQ} becomes 

\vspace{-5pt}
\begin{widetext}
\begin{alignat}{1}
\label{LLeq}
\textstyle
m_{\mbox{\tiny{obs}}}
\frac{\drm^2}{\drm\tau^2}\qQ
= 
 -\frac{e}{c}\FQ^{\mbox{\tiny{ext}}}(\qQ)\cdot\frac{\drm}{\drm\tau}\qQ 
 - \frac{2e^3}{3m_{\mbox{\tiny{obs}}}c^4} 
\left(\gQ +\frac{1}{c^2}\frac{\drm}{\drm\tau}\qQ \otimes\frac{\drm}{\drm\tau}\qQ\right)\cdot
\frac{\drm}{\drm\tau}
\left(\FQ^{\mbox{\tiny{ext}}}(\qQ)\cdot\frac{\drm}{\drm\tau}\qQ\right),
\end{alignat}
\end{widetext}
an implicit second-order ODE for the position of the electron, compatible with the available initial data.
 Equation \refeq{LLeq} is called the Eliezer--Ford--O'Connell equation in \cite{BurtonNoble}.
 The equation presented by {Landau--Lifshitz}~\cite{LandauLifshitz} differs from \refeq{LLeq} 
by an additional approximation: noting that
$\frac{\drm}{\drm\tau}
\left(\FQ^{\mbox{\tiny{ext}}}(\qQ)\cdot\frac{\drm}{\drm\tau}\qQ\right)
= 
\left(\frac{\drm}{\drm\tau}\FQ^{\mbox{\tiny{ext}}}(\qQ)\right)\cdot\frac{\drm}{\drm\tau}\qQ
+
\FQ^{\mbox{\tiny{ext}}}(\qQ)\cdot\frac{{\drm}^2}{\drm\tau^2}\qQ$, they
substitute $\frac{-e}{\mbare c}\FQ^{\mbox{\tiny{ext}}}(\qQ)\cdot\!\frac{\drm}{\drm\tau}\qQ$
for $\frac{{\drm}^2}{\drm\tau^2}\qQ$ in the last term.

 As recently as in \cite{PoissonETal} the  Eliezer--Ford--O'Connell equation \refeq{LLeq}, resp. its Landau--Lifshitz 
approximation, was still presented as the state of affairs
in the classical theory of point charge motion in flat spacetime.
 These equations owe their longevity to their reputation as \emph{practically effective equations of motion} for the 
computation of (first-order) radiation-reaction-corrected test-particle dynamics of point charges in smooth ``external'' fields.
 When point charges are replaced by extended charged particles the Landau--Lifshitz
equation can be derived rigorously from the Abraham--Lorentz model with nonzero bare mass, 
using center-manifold theory \cite{Spohn}, and also in a ``vanishing-particle limit'' \cite{GHW}.
 Its solutions are expected to agree reasonably well with empirical electron motion in the 
``classical regime'' of weak and slowly varying ``external'' fields,
and demands for higher precision can be met with improved effective equations, obtained either by adding higher-order classical
radiation-reaction correction terms, or by invoking QED.
 However, \refeq{LLeq} has major shortcomings even as an effective equation of motion for true point charges!

\vspace{-10pt}
                \subsection{Critique}\vspace{-5pt}

 To convey a first feeling for the limitations of the Eliezer--Ford--O'Connell equation \refeq{LLeq}
and its Landau--Lifshitz approximation, we recall the well-known fact that the last term in \refeq{LLeq} vanishes 
for electron motion along a constant applied electric field, and so does its Landau--Lifshitz approximation; cf. \cite{APMDb}.
 Thus in this textbook situation \refeq{LLeq} fails to take the energy-momentum loss due to radiation 
by the electron into account, i.e. its solution is identical to the familiar test-particle motion. 
 To radiation-reaction-correct these requires a non-vanishing higher-order term.

 A much more serious limitation of the Eliezer--Ford--O'Connell and 
Landau--Lifshitz equations was discovered recently \cite{DeckertHartenstein},
by considering the many-body version. 
 In this case each point charge satisfies its own equation \refeq{LLeq},
indexed by a subscript ${}_k$ (say) at $e_k$, $m_k$, $\qQ_k$, and at
$\FQ^{\mbox{\tiny{ext}}}_k$, where $\FQ^{\mbox{\tiny{ext}}}_k$ is now the Faraday tensor of the 
Maxwell--Lorentz field given by a sum of the Li\'enard--Wiechert fields \refeq{eq:LWsolE} and \refeq{eq:LWsolB}
of all the other particles but the $k$-th, plus the source-free field. 
 Since nobody knows the past histories of all the particles which enter the 
Li\'enard--Wiechert formulas \refeq{eq:LWsolE} and \refeq{eq:LWsolB}, one has to stipulate some past motions. 
 But whatever one stipulates, as shown in \cite{DeckertHartenstein}, \emph{typically} a singularity in the Maxwell--Lorentz
fields will propagate along the initial forward light-cone of each and every point charge, so that this system of 
equations of motion coupled with the Maxwell--Lorentz field equations is typically well-defined only until a point charge 
meets the forward initial light cone of another point charge.
 This is much too short a time span to be relevant to, e.g., plasma physics. 
 This problem cannot be overcome perturbatively by adding a higher-order radiation-reaction correction
term at r.h.s.\refeq{LLeq}. 
 Moreover, it's not just the radiation-reaction term in \refeq{LLeq} which causes trouble --- the expression of the
Lorentz force of one particle on another is typically not well-defined on the initial forward light-cones.
 Lorentz electrodynamics for $N>1$ point charges is in serious trouble!
 
 The above discussion leaves no room for reasonable doubts that ingenious extraction of effective equations of point charge motion 
from the mathematically ill-defined, merely symbolic ``Lorentz electrodynamics of point charges,''
is not a winning strategy to arrive at a mathematically well-defined and physically accurate relativistic theory of
point charge motion in the classical realm.
 In the remainder of this paper we explain how to formulate such a theory in a manner which 
 preserves the spirit of Lorentz electrodynamics as much as possible.

 To keep matters as simple as possible we first consider an electrodynamical system featuring only a single point charge.
 The $N$-body situation will be discussed in section \ref{sec:EMfN}.
 The motion in a constant applied electric field is revisited in section \ref{sec:EXAMPLES}.

\vspace{-10pt}
 
                \section{Basic definition of the electromagnetic force}\vspace{-5pt}\label{sec:force1}

 Mechanically the point charge is a point particle with a mechanical momentum 
\begin{equation}
\pV(t)
= \label{eq:MOMinTERMSofVELO}
\mbare\frac{\vV(t)}{\sqrt{1 -\frac{1}{c^2}|\vV(t)|^2}},
\end{equation}
with $\vV(t):=\Ddt \qV(t)$ its velocity and $\mbare\neq 0$ its \emph{bare rest mass}. 
  By  Newton's $2^{nd}$ \emph{law} the rate of change with time of the particle momentum equals the force acting on it, 
\begin{equation}
 \Ddt \pV(t)
= \label{eq:NewtonSECONDlaw}
\fV(t).
\end{equation}
 The force $\fV$ depends on the non-kinematical qualities of the point particle, in this case its electric charge which 
couples the point particle to the electromagnetic field.

 Electrodynamically the moving point charge is a source / sink for a classical electromagnetic field with 
\emph{field momentum (vector-)density} $\PiV^{\mbox{\tiny{field}}}(t,\sV)$. 
 Suppose the fields decay sufficiently rapidly as $|\sV|\to\infty$ so that $\PiV^{\mbox{\tiny{field}}}(t,\sV)$ is integrable 
w.r.t. $\drm^3s$ ``at spatial infinity.'' 
 Suppose also that the field singularity caused by the point charge is mild
enough so that $\PiV^{\mbox{\tiny{field}}}(t,\sV)$  is locally integrable over any neighborhood of $\qV(t)$, so that
 \begin{equation}
\pV^{\mbox{\tiny{field}}} (t) 
= \label{eq:FIELDp}
\int_{\Rset^3} \PiV^{\mbox{\tiny{field}}} (t,\sV) \drm^3s
\end{equation}
is a well-defined total field momentum vector. 
 Finally, suppose that the motion of the charge is sufficiently regular so that $\pV^{\mbox{\tiny{field}}} (t)$ is differentiable
with respect to time.

 Now, following Poincar\'e (cf. \cite{MillerBOOK}), we \emph{postulate} that in the absence of non-electromagnetic forces 
only those motions are permissible which satisfy the \emph{balance law}
\begin{equation}
\Ddt \pV(t) 
= \label{eq:dMOMbalance}
-  \Ddt \pV^{\mbox{\tiny{field}}} (t);
\end{equation}
i.e. any momentum gain by the particle is compensated through a corresponding momentum loss by the field, and
vice versa.
 It then follows from \refeq{eq:dMOMbalance} in concert with \refeq{eq:NewtonSECONDlaw} that 
{the electrodynamical force on a point charge source} of the classical electromagnetic field with a single source is to be defined by
\begin{equation}
\fV^{\mbox{\tiny{em}}}(t) 
:= \label{eq:EMf}
- \Ddt \int_{\Rset^3} \PiV^{\mbox{\tiny{field}}} (t,\sV) \drm^3s. 
\end{equation}
\begin{rem}
 Postulating the balance law \refeq{eq:dMOMbalance} is equivalent to postulating conservation of total momentum,
\begin{equation}
\Ddt \PV(t)
= \label{eq:TOTALpISconserved}
\NullV,
\end{equation}
for a total momentum defined as the sum of particle and field momenta, i.e.
\begin{equation}
\PV(t)
: = \label{eq:TOTALp}
 \pV(t) + \pV^{\mbox{\tiny{field}}} (t).
\end{equation}
\end{rem}

We pause for a moment to comment on \refeq{eq:EMf} in the context of Lorentz electrodynamics with point charges.

\vspace{-10pt}
                \subsection{Connection with Lorentz electrodynamics}\vspace{-5pt}

 If we assume the electromagnetic fields satisfy the Maxwell--Lorentz field equations with a point charge source
\refeq{eq:MLdotB} \&\ \refeq{eq:MLdotE}, and \refeq{eq:MLdivB} \&\ \refeq{eq:MLdivE}, in which case
$\PiV^{\mbox{\tiny{field}}} (t,\sV) = \frac{1}{4\pi c}\EV\crprd\BV$, then r.h.s.\refeq{eq:FIELDp} is generally ill-defined
(generally $\infty$ in magnitude), and then r.h.s.\refeq{eq:EMf} has no well-defined meaning either. 
 However, pretending that r.h.s.\refeq{eq:FIELDp} was well-defined, and that so was
r.h.s.\refeq{eq:EMf}, and furthermore pretending that all the ensuing (advanced) multi-variable calculus and analysis
steps were justified, r.h.s.\refeq{eq:EMf} would turn precisely into the expression of the Lorentz force \refeq{eq:Lf}.
 For later convenience we recall those steps: 

(a) pull the time derivative into the integral;

(b) apply the Leibniz rule to get $${\textstyle{\pddt \left(\EV\crprd\BV\right) =\left(\pddt\EV\right)\crprd\BV + \EV\crprd\pddt \BV}};$$

(c) now use \refeq{eq:MLdotB} to express the partial time derivative 

\ \ \ \ of $\BV$ in terms of $\nab\crprd\EV$, and \refeq{eq:MLdotE} to express the partial 

\ \ \ \ time derivatives of $\EV$ in terms of $\nab\crprd\BV$ and $\vV\delta_{\qV}$;

(d) use an advanced vector calculus identity in concert 

\ \ \ \ with \refeq{eq:MLdivB} \&\ \refeq{eq:MLdivE} to write the so manipulated r.h.s.\refeq{eq:EMf} 

\ \ \ \ as a volume integral over the sum of the divergence 

\ \ \ \  of Maxwell's stress tensor,  plus the Lorentz force 

\ \ \ \ vector density $-e\left(\EV+\frac1c\vV\crprd\BV\right)\delta_{\qV}$;

(e) use Gauss' theorem to conclude that the contribu-

\ \ \ \ tion from the stress tensor vanishes;

(f) carry out the volume integral of the force vector 

\ \ \ \ density  and thus obtain \refeq{eq:Lf}.

\noindent
 Remarkably, not a single one of these six steps is generally justified within the symbolic system of equations known as
``Lorentz electrodynamics with point charges.''
 Nevertheless this ``pseudo derivation'' of  \refeq{eq:Lf} from  \refeq{eq:EMf} does suggest that by replacing \refeq{eq:Lf} with
\refeq{eq:EMf} one may accomplish what was intended by Lorentz and his contemporaries.

 As a first encouraging observation we register that there is a large set of field initial data satisfying the constraint equations 
\refeq{eq:MLdivB} \&\ \refeq{eq:MLdivE} for which the formula \refeq{eq:EMf} of the electromagnetic force is \emph{initially} 
well-defined --- even for the Maxwell--Lorentz field theory! 
 In the special case of an \emph{electrostatic} field of a point charge at rest, the electromagnetic force \refeq{eq:EMf} is 
actually well-defined for all times and consistently equals $\NullV$, which it should in this case; cf. section 3.4.
 This already demonstrates the superiority of the formula \refeq{eq:EMf} over 
Lorentz' \refeq{eq:Lf}, which is ill-defined even in this simplest non-test charge situation.

 Unfortunately, replacing Lorentz' formula \refeq{eq:Lf} with \refeq{eq:EMf} does not convert Lorentz electrodynamics
with point charges into a well-defined theory:
there is an even larger set of Maxwell--Lorentz field initial data satisfying the constraint equations 
\refeq{eq:MLdivB} \&\ \refeq{eq:MLdivE} for which \refeq{eq:EMf} is not well-defined already at the initial time. 
 Also, even with favorable special Maxwell--Lorentz field initial data the electromagnetic initial force typically
cannot be continued into the future; i.e. typically the expression \refeq{eq:EMf} is not
well-defined for continuous stretches of time, the static special case being an exception. 
 Moreover, the initial energy density of an ML field with a point charge source is never locally integrable, i.e. 
for such data the field energy in any finite volume containing the point charge is infinite even if their field momentum 
\refeq{eq:FIELDp} exists.
 
\vspace{-15pt}
	\section{\hspace{-10pt}The role of the electromagnetic vacuum law}\label{sec:FIELDf}\vspace{-0.2truecm}

 In the following we will discuss \refeq{eq:EMf} for some electromagnetic field theories 
which (are expected to) yield a typically differentiable \refeq{eq:FIELDp}.
 Incidentally, \refeq{eq:EMf} was also the starting point of Abraham \cite{AbrahamD} and Lorentz \cite{LorentzENCYCLOP} 
for computing the electromagnetic force on a charged particle in their classical theories with extended electron models;
cf. \cite{JacksonBOOKb}.
 While Abraham and Lorentz and their peers chose to replace point electrons by extended structures but otherwise 
continued to work with the Maxwell--Lorentz field equations, we instead continue to work with point electrons but replace 
the Maxwell--Lorentz fields with solutions to the pre-metric Maxwell equations \cite{HehlObukov}  which satisfy different 
electromagnetic vacuum laws --- which furnish locally integrable energy-momentum densities of fields with point charge sources.
 (``Pre-metric'' means, the spacetime metric plays no role; see \cite{HehlObukov}. It enters through the vacuum law.)
 Since \emph{any classical electromagnetic field theory will be about some distinguished subset of solutions of 
the pre-metric Maxwell field equations}, the issue is indeed to identify the physically correct classical electromagnetic 
vacuum law!
 We refrain from trying to make such a definitive identification but instead consider two well-known proposals: the nonlinear 
system proposed by Born \&\ Infeld \cite{BornInfeldBb} (see also \cite{BiBiONE}), and the linear higher-order derivative system of
 Bopp \cite{BoppA,BoppB}, Land\'e \&\ Thomas \cite{Lande,LandeThomas}, and Podolsky \cite{Podolsky} (see also \cite{PodolskySchwed}).

\vspace{-10pt}
           \subsection{\hspace{-10pt}The pre-metric Maxwell field equations}\vspace{-5pt}\label{sec:preMeqns}
 The pre-metric Maxwell field equations are a four-dimensional (complex) analog of the familiar three-dimensional 
$\nab\cdot\BV = 0\implies \BV=\nab\crprd\AV$.
 Explicitly, the continuity equation 
\begin{equation}
\textstyle
        \pddt\rho (t,\sV) + \nab\cdot\jV(t,\sV)
=\label{eq:MrhojLAW}
	0 
\end{equation}
for the charge density $\rho$ and the current vector-density $\jV$ is a four-dimensional analog of $\nab\cdot\BV =0$. 
 It implies that $\rho$ and $\jV$ can be expressed as linear combination of first-order space and time derivatives of
two complex three-dimensional fields, $\DV+i\BV$ and $\HV-i\EV$; viz. 
$4\pi \rho = \nab\cdot (\DV+i\BV)$ and $4\pi\jV = c \nab\crprd (\HV-i\EV) - \pddt (\DV +i\BV)$.  
 The ``$4\pi$'' factor occurs for historical reasons, and the speed ``$c$'' at this point is just a conversion factor.
 Sorted into real and imaginary parts these are precisely the pre-metric Maxwell equations, which we write as one pair of 
homogeneous equations for $\BV$ and $\EV$, 
\begin{alignat}{1}
\textstyle
\pddt{\BV(t,\sV)} + c \nab\crprd\EV(t,\sV) 
&= \label{eq:MdotB}
\NullV \, , \\
        \nab\cdot \BV(t,\sV)  
&= \label{eq:MdivB}
        0\, ,
\end{alignat}
and one pair of inhomogeneous equations for $\DV$ and $\HV$,
\begin{alignat}{1}
\textstyle
 -\pddt{\DV(t,\sV)} + c\nab\crprd\HV(t,\sV)  
&=  
 4\pi  \jV(t,\sV) \,,
\label{eq:MdotD}\\
        \nab\cdot\DV(t,\sV)  
&=
        4 \pi \rho(t,\sV)\,. 
\label{eq:MdivD}
\end{alignat}
 Note that the constraint equations \refeq{eq:MdivB} \&\ \refeq{eq:MdivD} only impose on the initial data $\BV(0,\sV)$ and $\DV(0,\sV)$ 
which need to be supplied when viewing (as we will do) \refeq{eq:MdotB} \&\ \refeq{eq:MdotD} as initial value problems for $\BV$ and $\DV$,
respectively. 
 To see this for \refeq{eq:MdivB}, take the divergence of \refeq{eq:MdotB}; for
\refeq{eq:MdivD}, take the divergence of \refeq{eq:MdotD} and the time derivative of \refeq{eq:MdivD}, 
and recall \refeq{eq:MrhojLAW}.

 The pre-metric Maxwell equations are familiar from Maxwell's theory of electromagnetic fields in 
material media, though here they are used for fields sourced by point charges in an otherwise empty space.

\vspace{-10pt}
           \subsubsection{Their general solution for point charge sources}\label{sec:preMsolutions}\vspace{-5pt}
 The pre-metric Maxwell equations are easily solved if
the charge density $\rho(t,\sV) = -e \delta_{\qV(t)}(\sV)$ 
and the current vector-density $\jV(t,\sV)=  -e\delta_{\qV(t)}(\sV){\vV}(t)$,
provided that $t\mapsto\qV(t)$ is continuously differentiable so that the continuity equation \refeq{eq:MrhojLAW}
is automatically satisfied in the sense of distributions, and provided that $|\vV(t)|<c$.

 The pre-metric Maxwell equations in themselves can be viewed as two \emph{independent} systems of linear first-order PDE 
with constant coefficients, a homogeneous system for the field pair $(\BV,\EV)$, and an inhomogeneous system for the field 
pair $(\DV,\HV)$.
 Their general distributional solutions are readily written down, in the inhomogeneous case
conditioned on the motions of the point charges being given.
 For later convenience we collect the general solutions here; it suffices to do this for when 
there is only a single point charge. 

  The homogeneous system is solved by a linear combination of 
first-order derivatives of a vector potential field $\AV(t,\sV)\in \Rset^3$ and a scalar potential field $A(t,\sV)\in\Rset$, viz.
\begin{equation}\label{eq:BrepA}
\BV(t,\sV) = \nab \crprd \AV(t,\sV),
\end{equation}
\begin{equation}\label{eq:ErepA}
\EV(t,\sV) = -\nabla A(t,\sV) - \textstyle\pddct \AV(t,\sV).
\end{equation}
 Of course, this representation is found in every textbook on classical electrodynamics.
 
 Similarly we can handle the inhomogeneous equations. 
 Assuming the map $t\mapsto\qV(t)$ to be continuously differentiable, with a bounded Lipschitz continuous derivative $\vV(t)$
satisfying the speed limit $|\vV|<c$, 
the general solution to the system \refeq{eq:MdotD},  \refeq{eq:MdivD} then is
 \begin{equation}\label{eq:DrepZ}
 \DV(t,\sV) =  \DLWr(t,\sV)  + \nab\crprd \ZV(t,\sV),
\end{equation}
\begin{equation}\label{eq:HrepZ}
\HV(t,\sV) =  \HLWr(t,\sV)  + \nab Z(t,\sV) + \textstyle\pddct \ZV(t,\sV).
\end{equation}
 Here, $\DLWr$ and $\HLWr$ are the Li\'enard--Wiechert fields \refeq{eq:LWsolE} and \refeq{eq:LWsolB},
and the vector potential field $\ZV(t,\sV)\in \Rset^3$ and a scalar potential field $Z(t,\sV)\in\Rset$ generate
the general solution to the associated homogeneous system. 
 Note the sign difference between the homogeneous $(\BV,\EV)$ system and the homogeneous system
associated with the $(\DV,\HV)$ field pair.

 We remark that $\DV+i\BV = \DLWr + \nab\crprd (\ZV +i\AV)$ and that $\HV-i\EV = \HLWr + \nab(Z+iA) + \textstyle\pddct (\ZV+i\AV)$. 

\vspace{-5pt}
           \subsubsection{Gauge invariance}\label{sec:GAUGE}\vspace{-5pt}
 As is well-known, the r.h.s.s of \refeq{eq:BrepA}, \refeq{eq:ErepA} are invariant under the \emph{gauge transformation}
\begin{alignat}{1}\label{eq:gaugeT}
\AV(t,\sV)& \mapsto  \AV(t,\sV) + \nab \Upsilon(t,\sV) ,\\
A(t,\sV)  &\mapsto  A(t,\sV) - \textstyle \pddct \Upsilon(t,\sV) .
\end{alignat}
 Similarly,  the r.h.s.s of \refeq{eq:DrepZ}, \refeq{eq:HrepZ} are invariant under the \emph{gauge transformation}
\begin{alignat}{1}\label{eq:gaugeTagain}
\ZV(t,\sV) &\mapsto  \ZV(t,\sV) + \nab \mho(t,\sV) , \\ 
Z(t,\sV)  &\mapsto  Z(t,\sV) - \textstyle \pddct \mho(t,\sV) .
\end{alignat}

 The gauge transformations can be merged in complex notation: 
the addition of a four-dimensional ``pseudo gradient'' of a complex scalar $\mho +i\Upsilon$, i.e.
$(-\pddct,\nab)(\mho+i\Upsilon)$, to the complex four-dim. vector field $(Z+iA, \ZV+i\AV)$ does not change 
$\DV+i\BV$ and $\HV-i\EV$.

 \vspace{-10pt}

           \subsection{Electromagnetic vacuum laws}\label{sec:VACUUMlaws}\vspace{-5pt}
 The Maxwell--Lorentz field theory is concerned exclusively with those solutions of the
pre-metric Maxwell field equations whose imaginary and real parts (referring to the 
fields $\DV+i\BV$ and $\HV-i\EV$) are related by
\begin{alignat}{1}
        \HV(t,\sV)  
&= \label{eq:MlawBisH}
        \BV(t,\sV)  \, ,
\\
        \DV(t,\sV)  
&=
        \EV(t,\sV) \, ,
\label{eq:MlawEisD}
\end{alignat}
which crosslinks the homogeneous with the inhomogeneous pair of equations.
 Equations \refeq{eq:MlawBisH} \&\ \refeq{eq:MlawEisD} are known as Maxwell's 
law of the electromagnetic vacuum (``law of the pure ether'' in Maxwell's words).
 As explained in the introduction, Maxwell's law of the electromagnetic vacuum selects solutions of the pre-metric
Maxwell field equations which are too singular to allow a well-defined coupling with the classical (relativistic or not)
theory of point particle motion. 
 But there are more suitable electromagnetic vacuum laws which express the real parts 
of $\DV+i\BV$ and $\HV-i\EV$ in terms of the imaginary parts. 

 As shown by Mie \cite{MieFELDTHEORIEa}, \cite{MieFELDTHEORIEb}, in a Lorentz co-variant electrodynamics the vacuum law follows
from a Lorentz-scalar Lagrangian (density) $\cL$. 
 The notion of Lorentz invariance makes it obvious that the spacetime metric enters at this point. 
 In the orthodox version $\cL$ depends only on the Lorentz invariants $|\EV|^2-|\BV|^2$ and $(\EV\cdot\BV)^2$, but
Lagrangians which in addition depend on the Lorentz invariant $(\nab\cdot\EV)^2 - |\nab\crprd\BV-\frac1c\pddt\EV|^2$
have also been considered in the literature (see below).
 The fields $\DV$ and $\HV$ are in either case obtained by functional differentiation from the action $\cA=\int \cL \drm^3{s}\drm{t}$, 
viz. $\DV = \delta_{\EV} \cA$ and $\HV = - \delta_{\BV}\cA$. 
 If $\cL$ depends only on the invariants  $|\EV|^2-|\BV|^2$ and $(\EV\cdot\BV)^2$, this is equivalent to conventional
partial differentiation of the Lagrangian density, viz. $\DV = \partial_{\EV} \cL$ and $\HV = - \partial_{\BV}\cL$.

 We next list the field Lagrangians and the implied electromagnetic vacuum laws for the ML, the MBI, and the MBLTP
field equations,  in historical order.

           \subsubsection{Schwarzschild's field Lagrangian and Maxwell's vacuum law}\vspace{-5pt}\label{sec:MLlaw}
%
 Schwarzschild's \cite{Schwarzschild} Lagrangian, given by
\begin{alignat}{1}
 4\pi \cL_{\mbox{\tiny{S}}}
= \label{eq:SchwaL}
\tfrac{1}{2} \left(|\EV|^2-|\BV|^2\right),
\end{alignat}
yields  Maxwell's ``law of the pure ether,'' \refeq{eq:MlawBisH} \&\ \refeq{eq:MlawEisD}, obeyed by
the Maxwell--Lorentz fields.
 We already reviewed the Maxwell--Lorentz field equations in the introduction. 

 Even though \refeq{eq:SchwaL} is not an admissible field Lagrangian for a classical electrodynamics 
with point charges, the success of the Maxwell--Lorentz field equations in the realm of \emph{weak-field} phenomena 
(i.e. far away from the hypothetical point charge sources) and the \emph{low-frequency / long wavelength} regime
(i.e. visible light, infra-red, radio waves and such) suggests that 
every admissible Lagrangian must reduce to it in the weak-field and low-frequency / long wavelength regime.

           \subsubsection{The Born--Infeld field Lagrangian and vacuum law}\vspace{-5pt}\label{sec:BIlaw}
%
 The Born--Infeld field Lagrangian \cite{BornInfeldBb}, given by
\begin{alignat}{1}
 4\pi \cL_{\mbox{\tiny{BI}}}
= \label{eq:BIL}
b^2 - \sqrt{b^4 - b^2\left(  |\EV|^2-|\BV|^2\right) - \left(\EV\cdot\BV\right)^2 }
\end{alignat}
yields the Born--Infeld (BI) law of the electromagnetic vacuum, 
\begin{alignat}{1}
        \HV
&= \label{eq:BIlawHofEB}
\frac{\BV - \frac{1}{b^2}(\BV\cdot\EV)\,\EV}{\sqrt{1-\frac{1}{b^2}(|\EV|^2-|\BV|^2) - \frac{1}{b^4}(\EV\cdot\BV)^2}} \, , \\
        \DV
&=
\frac{\EV + \frac{1}{b^2}(\BV\cdot\EV)\,\BV}{\sqrt{1-\frac{1}{b^2}(|\EV|^2-|\BV|^2) - \frac{1}{b^4}(\EV\cdot\BV)^2}}\, ,
\label{eq:BIlawDofEB}
\end{alignat}
expressing the pair $(\DV,\HV)$ in terms of the pair $(\BV,\EV)$.
 The parameter $b$ is Born's field strength constant.
 In the limit $b\to\infty$ the BI law converges to Maxwell's law. 

 Since mathematically the pre-metric Maxwell field equations are quite
naturally interpreted as a pair of evolutionary equations \refeq{eq:MdotB} \&\ \refeq{eq:MdotD}
for the fields $\BV$ and $\DV$, with initial data which are
constrained by \refeq{eq:MdivB} \&\ \refeq{eq:MdivD}, it is desirable to rather express the field pair 
$(\EV,\HV)$ in terms of the pair $(\BV,\DV)$. 
 Happily \refeq{eq:BIlawHofEB}, \refeq{eq:BIlawDofEB} can be converted into
\begin{alignat}{1}
&\HV=\frac{\BV - \frac{1}{b^2}\DV\times(\DV\times\BV)}{\sqrt{1+\frac{1}{b^2}(|\BV|^2+|\DV|^2) +\frac{1}{b^4}|\BV\times\DV|^2}}\,,
\label{eq:BIlawHofBD}
 \\
&\EV=\frac{\DV - \frac{1}{b^2}\BV\times(\BV\times\DV)}{\sqrt{1+\frac{1}{b^2}(|\BV|^2+|\DV|^2) +\frac{1}{b^4}|\BV\times\DV|^2}}\,.
\label{eq:BIlawEofBD}
\end{alignat}
  Complemented with \refeq{eq:BIlawHofBD} \&\ \refeq{eq:BIlawEofBD} the pre-metric Maxwell field equations \refeq{eq:MdotB} \&\ \refeq{eq:MdotD}
turn into the Maxwell--Born--Infeld (MBI) evolution equations for the fields $\BV$ and $\DV$, their initial data being constrained by
\refeq{eq:MdivB} \&\ \refeq{eq:MdivD}. 

 In the absence of any sources the initial value problem for the MBI field equations is globally well-posed for 
classical initial data with sufficiently small energy \cite{SpeckMBI}. 
 However, it is also known that certain smooth plain-wave data can lead to a singularity after a finite time, see
\cite{SerreA,Brenier}.
 Speck in his thesis showed that this can be extended to finite energy data which coincide with such plane wave data on 
a sufficiently large bounded domain in space. 
 It is not known whether such blow-up in finite time happens for all finite-energy data in an open neighborhood 
of these finite-energy ``local plane-wave'' type data.

 For the MBI field equations with fixed point charge sources it has been shown \cite{KieMBIinCMP} that a unique 
finite energy electrostatic solution exists which is real analytic except at the locations of the point charges; this
holds for any finite number $N$ of point charges with arbitrary signs, magnitudes, and placements.
 However, it is not yet known whether the nonlinear BI law leads to an at least locally well-posed 
initial value problem for a physically interesting class of Maxwell--Born--Infeld fields with point charge sources.
 
\vspace{-5pt}
           \subsubsection{Bopp's field Lagrangian and vacuum law}\vspace{-5pt}\label{sec:BLTPlaw}
%
 In the 1940s  Bopp, Land\'e \&\ Thomas, and Podolsky argued that a more accessible \emph{linear} vacuum law is available
if one is willing to admit higher-order derivative electromagnetic field equations.
 Bopp \cite{BoppA} obtained the equations from a Lagrangian given by
\begin{alignat}{1}
 4\pi \cL_{\mbox{\tiny{BLTP}}}
= \label{eq:BLTPL}
 &\tfrac12\left(|\EV|^2-|\BV|^2\right) \\ \notag
 & + \tfrac12 \tfrac{1}{\varkappa^2} \Big[(\nab\cdot\EV)^2 - |\nab\crprd\BV-\tfrac1c\textstyle\pddt\EV|^2\Big]
\end{alignat}
which yields the Bopp--Land\'e--Thomas--Podolsky (BLTP) electromagnetic vacuum law\vspace{-3pt}
\begin{alignat}{1}
        \HV(t,\sV)  
&= \label{eq:BLTPlawBandH}
       \left(1  + \varkappa^{-2}\square\,\right) \BV(t,\sV) \, ,
\\
        \DV(t,\sV) 
&=
        \left(1  + \varkappa^{-2}\square\,\right) \EV(t,\sV) \, ;
\label{eq:BLTPlawEandD}
\end{alignat}
here, $\square \equiv c^{-2}\partial_t^2 -\Delta$  is the classical wave operator.
 The parameter $\varkappa$ is ``Bopp's reciprocal length'' \cite{BoppA}; see \cite{CKP} for empirical constraints on $\varkappa$.
 The singular limit $\varkappa\to\infty$ of the BLTP law yields Maxwell's law. 

 The pre-metric Maxwell field equations \refeq{eq:MdotB} \&\ \refeq{eq:MdotD} and \refeq{eq:MdivB} \&\ \refeq{eq:MdivD},
when supplemented by the BLTP law of the vacuum \refeq{eq:BLTPlawBandH} \&\ \refeq{eq:BLTPlawEandD}, become the
Maxwell--Bopp--Land\'e--Thomas--Podolsky (MBLTP) field equations. 
 Different from the Maxwell--Lorentz and Maxwell--Born--Infeld field equations, they are higher-order derivative field 
equations, requiring initial data not only for $\BV$ and $\DV$, but in addition also for $\EV$ and $\partial_t\EV =: \dot\EV$.

 We pause for another moment and comment on the asymmetrical role played by the pair of equations 
\refeq{eq:BLTPlawBandH} \&\ \refeq{eq:BLTPlawEandD}, despite their symmetric appearance.
 When judged in their own right, 
\refeq{eq:BLTPlawEandD} is a second-order evolution equation for the electric field $\EV$, given $\DV$, and 
\refeq{eq:BLTPlawBandH} is a second-order evolution equation for $\BV$, given $\HV$. 
 However, since \refeq{eq:BLTPlawBandH} and \refeq{eq:BLTPlawEandD} are coupled with 
the pre-metric Maxwell evolution equations \refeq{eq:MdotB} \&\ \refeq{eq:MdotD} for the
fields $\BV$ and $\DV$ (constrained by \refeq{eq:MdivB} \&\ \refeq{eq:MdivD}), appearances are misleading
in the case of \refeq{eq:BLTPlawBandH}.
 A well-defined initial value problem for the fields is obtained only if \refeq{eq:MdotB} \&\ \refeq{eq:MdotD} 
and \refeq{eq:BLTPlawEandD} are treated as genuine evolution equations for the fields $\BV$, $\DV$, and $\EV$, while 
\refeq{eq:BLTPlawBandH} is not treated as an evolution equation for $\BV$ --- against all appearances.

 Indeed, given the field initial data $\EV(0,\sV)$, also $\nab\crprd\EV(0,\sV)$ is fixed initially, so \refeq{eq:MdotB} 
yields $\dot\BV(0,\sV)$.
 And given the field initial data $\dot\EV(0,\sV)$, also $\nab\crprd\dot\EV(0,\sV)$ is  fixed initially, so 
the time derivative of \refeq{eq:MdotB} yields $\ddot\BV(0,\sV)$. 
 And with the initial data $\BV(0,\sV)$ given, also $\Delta \BV(0,\sV)$ is fixed, and then 
r.h.s.\refeq{eq:BLTPlawBandH} is completely determined initially.  
  Thus \refeq{eq:BLTPlawBandH} \emph{defines} $\HV$ initially in terms of $\BV$ and its second partial derivatives; 
note that this also implies that $\nab\cdot\HV=0$. 
 And then, with $\HV$ so defined initially, and the particle's initial position and velocity given,
\refeq{eq:MdotD} now yields $\dot\DV(0,\sV)$. 
 Lastly, with the initial field data $(\EV, \dot\EV)(0,\sV)$, and $\DV(0,\sV)$ given, 
$\ddot\EV(0,\sV)$ is initially determined by \refeq{eq:BLTPlawEandD}.

 This scheme now propagates in time, i.e. \refeq{eq:BLTPlawBandH} remains the defining equation for $\HV$ 
in terms of $\BV$ and its second partial derivatives, while \refeq{eq:MdotB},  \refeq{eq:MdotD}, and \refeq{eq:BLTPlawEandD}
are genuine evolution equations for $\BV$, $\DV$, and $\EV$. 
 In \cite{KTZonBLTP} it is shown that MBLTP field initial data 
$(\BV, \DV, \EV, \dot\EV)(0,\sV)$ launch a unique global distributional solution of the MBLTP field equations,
conditioned on the motions being given.

 There is a small variation on this theme, which takes advantage of the convenience of having the general distributional
solution of the pre-metric Maxwell field equations for the pair $(\DV,\HV)$ available with \refeq{eq:DrepZ} \&\ \refeq{eq:HrepZ}.
 Thus, prescribing the motion $t\mapsto\qV(t)$ for $t\leq 0$ conveniently, though twice
continuously differentiable with subluminal velocity $\vV(t)$, and choosing smooth and spatially rapidly decaying fields
$Z(0,\sV)$, $\ZV(0,\sV)$, and $(\pddt\ZV)(0,\sV)$, such that $\nab\cdot\HV=0$, equations \refeq{eq:DrepZ} \&\ \refeq{eq:HrepZ} 
fix $\DV(0,\sV)$ and $\HV(0,\sV)$. 
 Prescribing also $(\EV, \dot\EV)(0,\sV)$ fixes $\nab\crprd\EV(0,\sV)$ and $\nab\crprd\dot\EV(0,\sV)$ initially, so
\refeq{eq:MdotB} and its time derivative yield $(\pddt\BV)(0,\sV)$ and $(\pddtSQUARE\BV)(0,\sV)$. 
 Thus, \refeq{eq:BLTPlawBandH}, while still not an evolution equation for $\BV$, is now an elliptic vector Helmholtz equation for $\BV(0,\sV)$,
which has a unique solution that vanishes at spatial infinity, thus determining the initial $\BV(0,\sV)$ completely.

 We next collect the pertinent formulas for the field momentum vector-densities, then show that there are 
field initial data, satisfying the Maxwell constraint equations, for which \refeq{eq:EMf} is initially well-defined.
 Lastly we address the electrodynamical admissibility of the vacuum laws.

	\subsection{The electromagnetic field energy-momentum density}\label{sec:FIELDp}\vspace{-0.2truecm}
 A Lorentz invariant field Lagrangian also determines the field energy-momentum density.
 For the Maxwell--Lorentz (ML) field theory, the field energy density $\veps^{\mbox{\tiny{ML}}}$ and 
{field momentum vector-density} $\PiV^{\mbox{\tiny{ML}}}$ are of course well known and 
given by
\begin{equation}
4\pi \veps^{\mbox{\tiny{ML}}}
= \label{eq:TooML}
  \tfrac12\big(|\BV|^2 +|\DV|^2\big) \,,
\end{equation}
\begin{equation}
\textstyle
4\pi c \PiV^{\mbox{\tiny{ML}}}
= \label{eq:PiML}
\DV\crprd\BV \,.
\end{equation}
  Recall that $\DV=\EV$ in Maxwell--Lorentz field theory.

 For the Maxwell--Born--Infeld (MBI) field theory, the field energy density $\veps^{\mbox{\tiny{MBI}}}$ and 
{field momentum vector-density} $\PiV^{\mbox{\tiny{MBI}}}$ are given by
\begin{equation}
4\pi \veps^{\mbox{\tiny{MBI}}}
= \label{eq:TooMBI}
{\sqrt{{b^4}+{b^2}(|\BV|^2+|\DV|^2) + |\BV\times\DV|^2}} -b^2 \,,
\end{equation}
\begin{equation}
\textstyle
4\pi c \PiV^{\mbox{\tiny{MBI}}}
= \label{eq:PiMBI}
\DV\crprd\BV .
\end{equation}

 For the Maxwell--Bopp--Land\'e--Thomas--Podolsky (MBLTP) field theory, the field energy density $\veps^{\mbox{\tiny{MBLTP}}}$ and 
{field momentum vector-density} $\PiV^{\mbox{\tiny{MBLTP}}}$ are given by
\begin{alignat}{1}
4\pi \veps^{\mbox{\tiny{MBLTP}}}
= \label{eq:TooMBLTP}
 & \BV\cdot\HV +  \EV\cdot\DV  - \tfrac12\big(|\BV|^2 +|\EV|^2\big) \\ \notag
& -  \tfrac{1}{2\varkappa^2}  \Big[\big(\nabla\cdot\EV\big)^2 + \big|\nabla\crprd\BV - {\textstyle\frac1c\pddt}\EV\big|^2 \Big]\,,
\end{alignat}
\begin{alignat}{1}
\textstyle
4\pi c \PiV^{\mbox{\tiny{MBLTP}}}
= \label{eq:PiMBLTP}
 & \DV\crprd\BV + \EV\crprd\HV - \EV\crprd\BV \\ \notag
& -  \tfrac{1}{\varkappa^2}  \big(\nabla\cdot\EV\big)\Big(\nabla\crprd\BV - \textstyle\frac{1}{c}\pddt\EV\Big).
\end{alignat}

	\subsection{Field data yielding an electromagnetic force initially}\label{sec:FIELDfINIT}\vspace{-0.2truecm}
  Consider first the ML and MBI field theories. 
 Both operate with the same formula for the field momentum density, \refeq{eq:PiML} respectively \refeq{eq:PiMBI}. 
 Initial data for the fields $\BV$ and $\DV$ compatible with the constraint equations \refeq{eq:MdivB} \&\ \refeq{eq:MdivD} 
for which \refeq{eq:FIELDp} is initially well-defined are easily obtained as follows.

 Set $\DV(0,\sV) = e\nab\frac{1}{|\sV-\sV(0)|} - \nab Z(0,\sV) -(\pddct\ZV)(0,\sV)$ and $\BV(0,\sV) = \nab\crprd\AV(0,\sV)$,
with $Z(0,\sV)$ and $(\pddct\ZV)(0,\sV)$ and $\AV(0,\sV)$ smooth and rapidly decaying at spatial infinity together with
their derivatives. 
 Then $(\DV\crprd\BV)(0,\sV)$ is integrable over $\Rset^3$, i.e. $\pV^{\mbox{\tiny{field}}} (0)$ exists.
 
 As to the derivative of $\pV^{\mbox{\tiny{field}}} (t)$ at $t=0$, consider first the ML field theoy. 
 With the above choice of initial data, also $\nab\crprd \DV(0,\sV)$ is smooth and so is $\nab\crprd \BV(0,\sV)$.
 Given in addition the assumed decay at spatial $\infty$, 
steps (a), (b), and (c) from section II.A are now justified to manipulate \refeq{eq:EMf} and yield
\begin{widetext}
\begin{equation}
\fV^{\mbox{\tiny{em}}}(0) = \label{eq:EMfMLinit}
  \int_{\Rset^3} \left[\BV\crprd\nab\crprd\BV + \DV\crprd\nab\crprd\DV +4\pi\tfrac1c\jV\crprd\BV\right](0,\sV) \drm^3s,
\end{equation}
\end{widetext}
which is well-defined.
 Note though that $(\rho\DV)(0,\sV)$ is not well-defined, so that one cannot apply steps (d) and (e)
of section II.A and arrive at the Lorentz formula for $\fV^{\mbox{\tiny{em}}}(0)$.

 Consider next the MBI field theory.
 If we formally carry out steps (a), (b), and the analog of (c) of section II.A, we get
\begin{widetext}
\begin{equation}
\fV^{\mbox{\tiny{em}}}(0) = \label{eq:EMfMBIinitA}
  \int_{\Rset^3} \left[\BV\crprd\nab\crprd\HV + \DV\crprd\nab\crprd\EV +4\pi\tfrac1c\jV\crprd\BV\right](0,\sV) \drm^3s,
\end{equation}
\end{widetext}
which may or may not be well-defined, depending on $\BV(0,\sV)$.
 Clearly the $\jV\crprd\BV$ term is the same as in the ML setup, pairing a $\delta$ distribution with a smooth test function.
 The $\DV\crprd\nab\crprd\EV$ term is also well-defined, and integrable, because the BI law \refeq{eq:BIlawEofBD} guarantees 
(i) that $|\EV|$ is uniformly bounded whenever $|\BV|$ is, and (ii) that for our initial data $\EV(0,\sV)$ is differentiable 
everywhere except at $\sV=\qV(0)$, having uniformly bounded partial derivatives which (iii) decay not slower than $\frac{1}{r^2}$ 
at spatial infinity.
 However, unless $\BV(0,\sV)$ vanishes at $\sV=\qV(0)$, 
\refeq{eq:BIlawHofBD} yields that $\HV(0,\sV)$ has a $\frac{1}{r^2}$ singularity at $\sV=\qV(0)$, 
and it is easy to check that its curl 
will then generally blow up like $\frac{1}{r^3}$, which is too strongly.
 If $\BV(0,\sV)$ vanishes like $|\sV-\qV(0)|$ or faster at $\sV=\qV(0)$, then
the $\BV\crprd\nab\crprd\HV$ term is well-defined, and integrable.
 Note that \refeq{eq:Lf} is still not well-defined even
with such idealized initial data because $\EV(0,\sV)$ is not well-definable at $\sV=\qV(0)$.

 Since typically the $(\BV\crprd\nab\crprd\HV)(0,\sV)$ term is not absolutely integrable over a 
neighborhood of $\sV=\qV(0)$, one typically is not allowed to carry out the manipulations by means 
of which one arrives at \refeq{eq:EMfMBIinitA}.
 Nevertheless, since \refeq{eq:PiMBI} has at most $1/r^2$ singularities, 
 \refeq{eq:EMf} may still be well-defined for MBI fields with a regularly moving point charge.
 This is yet to be sorted out, though.

 Lastly, as to the MBLTP field theory, it is not necessary to separately discuss the possibility of an initially well-defined 
force because below we will report that we actually were able to obtain a generically well-defined force which persists into 
the future. 

\vspace{-10pt}
	\subsection{\hspace{-0.2truecm}Electrodynamic admissibility of the vacuum laws}\label{sec:EMfGENERIC}\vspace{-0.2truecm}
 The examples discussed in the previous subsection demonstrate that the definition \refeq{eq:EMf} can yield a 
well-defined expression for the electromagnetic force on a point charge source of certain classical electromagnetic fields,
whereas the Lorentz formula \refeq{eq:Lf} fails to be well-defined.
 This is encouraging but does not suffice to accomplish our goal, which is to show that a generically well-defined, in fact 
well-posed and physically interesting classical electrodynamics with point charges is possible.
 Our next step thus is to inquire into field evolutions for which  \refeq{eq:EMf} remains well-defined over time.

	\subsubsection{The ML vacuum law is electrodynamically inadmissible}\label{sec:MLf}\vspace{-0.2truecm}
 For the sake of completeness, we note that the electrostatic special case
leads to a consistent proper solution of Lorentz electrodynamics in which the Maxwell--Lorentz field equations are coupled
with \refeq{eq:EMf} instead of \refeq{eq:Lf}.
 Indeed, assuming that $\qV(t)=\qV(0)$ and $\vV(t)=\NullV$ for all time, field initial data 
$\DV(0,\sV)= e\nab\frac{1}{|\sV-\sV(0)|}$ and $\BV(0,\sV) = \NullV$ propagate in time unchanged, viz. 
$\DV(t,\sV)= \DV(0,\sV)$ and $\BV(t,\sV) = \NullV$ for all time. 
 But then $(\DV\crprd\BV)(t,\sV) = \NullV$ for all time, and so  by \refeq{eq:EMf} also $\fV^{\mbox{\tiny{em}}}(t)=\NullV$ for all time, 
consistent with $\vV(t)=\NullV$ and $\qV(t)=\qV(0)$ for all time. 

 Unfortunately this very special situation does not have an open dynamical neighborhood in Lorentz electrodynamics even when
\refeq{eq:Lf} is replaced by \refeq{eq:EMf}.
 Suppose we perturb the static field data $\DV(0,\sV)= e\nab\frac{1}{|\sV-\sV(0)|}$ and $\BV(0,\sV) = \NullV$ 
by replacing the vanishing magnetic initial induction by a smooth and compactly supported $\BV(0,\sV) = \nab\crprd\AV(0,\sV)$ 
such that  $\BV(0,\sV)\crprd\nab\crprd \BV(0,\sV)$ is integrable over $\Rset^3$. 
 Even leaving the particle initial data as before, by \refeq{eq:EMfMLinit} the point charge will now feel a nonzero
initial force and begin to move. 
 But then by the Maxwell--Lorentz field equations the magnetic induction will \emph{not}
evolve into a smooth $\BV(t,\sV)$; a $\frac{1}{r^2}$ singularity is formed after an arbitrarily short time span after the 
initial instant, so $\DV\crprd\BV$ is only integrable at $t=0$, and the force  \refeq{eq:EMf} does not remain
well-defined. 
 This disqualifies the Maxwell--Lorentz field equations.

	\subsubsection{The BI vacuum law may be electrodynamically admissible}\label{sec:MBIf}\vspace{-0.2truecm}
 Everything we wrote about the electrostatic special case in Lorentz electrodynamics, with \refeq{eq:Lf} replaced by
\refeq{eq:EMf}, carries over to Born--Infeld electrodynamics. 
 However, this time there may be an open neighborhood of initial data for which 
the ensuing MBI field evolution with a large open set of assumed motions of their point source leads to a well-defined
\refeq{eq:EMf} as time goes on.
  To be sure, no such result has been proven rigorously yet, but 
the author is optimistic that the BI law is admissible, i.e. that an open set of solutions to the MBI field equations 
with point charge sources has integrable field momentum densities, and that the total field momentum is differentiable 
in time --- for not too long time intervals. 

	\subsubsection{The BLTP vacuum law is electrodynamically admissible}\label{sec:MBLTPf}\vspace{-0.2truecm}
 In \cite{KTZonBLTP} we establish the electrodynamic admissibility of the BLTP law, which here we summarize for
the simplified case of fields with a single point charge source. 
 We begin with a definition.
\begin{dfn}\label{dfn:BLTPfieldINITIALdata}
 Electrodynamically admissible initial data are of the following form: $(\BV, \DV, \EV, \dot\EV)(0,\sV)$
$=\sum_{j=0}^1(\BV_j, \DV_j, \EV_j, \dot\EV_j)(0,\sV)$. 
 Here, $(\BV_0, \DV_0, \EV_0, \dot\EV_0)(0,\sV)$ is the $t=0$ evaluation of a $C^{0,1}$ finite-energy source-free 
MBLTP solution which is globally bounded by $(\mathcal{B}_0, \mathcal{D}_0, \mathcal{E}_0, \dot{\mathcal{E}}_0)$, 
having global Lipschitz constants $L_{\BV_0}$ and $L_{\EV_0}$.
 Moreover, $(\BV_1, \DV_1, \EV_1, \dot\EV_1)(0,\sV)$ is the ``co-moving electromagnetic field''  at $t=0$ of a 
fictitious point charge whose world line coincides with the tangent world line of the actual point charge at $t=0$.
\end{dfn}
\vspace{-5pt}
\begin{rem}
 Replacing the ``co-moving electromagnetic field,'' of a fictitious point charge whose world line 
coincides with the tangent world line of the actual point charge at $t=0$, by the retarded Li\'{e}nard--Wiechert-type 
fields of a fictitious point charge whose subluminal $C^{1,1}$ world line merely is tangent to the world line of the 
actual point charge at $t=0$, does \emph{not} result in more general initial data, then, because the difference 
of two such fields at time $t=0$ is regular enough and satisfies the source-free MBLTP field equations at $t=0$. 
\end{rem}

 The proof in  \cite{KTZonBLTP} of the electrodynamical admissibility of the BLTP law of the electromagnetic vacuum 
consists in showing that the electrodynamically admissible initial data stipulated above launch field evolutions for 
which the force \refeq{eq:EMf} is well-defined for an open set of physically acceptable motions. 
 This proof is greatly facilitated by the fact that the electromagnetic force in the BLTP vacuum can be computed 
explicitly; the details are given in \cite{KTZonBLTP}.
 For the convenience of the reader the result of this computation is summarized next; 
it has also been announced in a conference proceedings, see \cite{KTZonEIHa}.

	\section{The electromagnetic force in the BLTP vacuum}\label{sec:BLTPf}\vspace{-0.2truecm}
 By the linearity of the MBLTP field equations the solution launched by the above stated type of initial data
decomposes into the pertinent sum of a vacuum field plus a Li\'enard--Wiechert(-type) field.
 The vacuum field need not be represented explicitly for it suffices to know that it has the required regularity.
 The field solutions $\DV_1(t,\sV)$ and $\HV_1(t,\sV)$ for $t> 0$ are for a.e. $|\sV-\qV(t)|>0$
given by \emph{the} Li\'enard--Wiechert fields \refeq{eq:LWsolE} and \refeq{eq:LWsolB}, 
\vskip-.6truecm
\begin{alignat}{1}
{\DV_1(t,\sV)} &=\label{eq:LWsolD} \DLWr(t,\sV)\, ,
\\
\hskip-1truecm
{\HV_1(t,\sV)} 
&= \label{eq:LWsolH} \HLWr(t,\sV)
\, .\vspace{-10pt}
\end{alignat}
 The MBLTP field solutions $\BV_1(t,\sV)$ and $\EV_1(t,\sV)$ for $t\geq 0$ are given by (cf. \cite{KTZonBLTP})
\begin{widetext}
\begin{alignat}{1}
\hspace{-20pt}
{\EV_1(t,\sV)} =\; \label{eq:EjsolMBLTP} 
&
 -e^{}\varkappa^2\tfrac12\tfrac{\nV(\qV^{},\sV)-{\vV^{}}/{c}}{1-\nV(\qV^{},\sV)\cdot{\vV}/{c}}
\Big|_{\mathrm{ret}}
\, + 
e^{} \varkappa^2 \int_{-\infty}^{t^\mathrm{ret}(t,\sV)}
\tfrac{J_2\!\bigl(\varkappa\sqrt{c^2(t-t')^2-|\sV-\qV(t')|^2}\bigr)}{{c^2(t-t')^2-|\sV-\qV(t')|^2}^{\phantom{n}} }
c \left(\sV-\qV^{}(t') - \vV^{}(t')(t-t')\right)\drm{t'}, \\ 
\hspace{-20pt}
{\BV_1(t,\sV)} =\; \label{eq:BjsolMBLTP}
& -e^{} \varkappa^2 \tfrac{1}{2} 
\tfrac{{\color{black}\vV^{}\crprd\nV(\qV^{},\sV)/c}}{1-\nV(\qV^{},\sV)\cdot{\vV}/{c}}
\Big|_{\mathrm{ret}}
\, + 
 e^{} \varkappa^2 \int_{-\infty}^{t^\mathrm{ret}(t,\sV)}
\tfrac{J_2\!\bigl(\varkappa\sqrt{c^2(t-t')^2-|\sV-\qV(t')|^2 }\bigr)}{{c^2(t-t')^2-|\sV-\qV(t')|^2}^{\phantom{n}} }
{\vV^{}(t')}\crprd \left(\sV-\qV^{}(t') 
\right)\drm{t'} ;
\end{alignat}
\end{widetext}
note that the time integrations from $-\infty$ to $0$ here do \emph{not} involve some unknown past motion but only the auxiliary straight-line 
motion which encodes the Lorentz-boosted electrostatic MBLTP field of the point charge, to which expressions \refeq{eq:EjsolMBLTP}
and \refeq{eq:BjsolMBLTP} reduce at $t=0$.

 With the help of these solution formulas the electromagnetic force of the MBLTP field on its point charge source
can be computed as follows. 
 Since each electromagnetic field component is the sum of a vacuum field and a sourced field, the bilinear $\PiV^{\mbox{\tiny{MBLTP}}}$
decomposes into a sum of three types of terms, the vacuum-vacuum terms, the source-source terms, and the mixed vacuum-source terms.
 The vacuum-vacuum contribution to r.h.s.\refeq{eq:EMf} vanishes because the total momentum of a vacuum field is conserved;
 the vacuum-source contribution to r.h.s.\refeq{eq:EMf} yields the force on the point source due to the vacuum field;
 lastly, the source-source contribution to r.h.s.\refeq{eq:EMf} is a ``self''-field force in BLTP electrodynamics. 
 Thus, \refeq{eq:EMf} is given by
\begin{equation}\label{eq:totalF}
\fV^{\mbox{\tiny{em}}}(t) 
=
\fV^{\mbox{\tiny{vacuum}}}[\qV,\vV](t)
+
\fV^{\mbox{\tiny{source}}}[\qV,\vV;{\color{blue}\aV}](t)
\end{equation} 
where
\begin{equation}\label{eq:vacuumF}
\fV^{\mbox{\tiny{vacuum}}}[\qV,\vV](t) =
- e \left[ \EV_0(t,\qV(t))+ {\color{red}\tfrac1c}\vV(t)\times \BV_0(t,\qV(t))  \right]
\end{equation} 
is the Lorentz force \refeq{eq:Lf} evaluated with a vacuum field (i.e. a ``test particle contribution'' 
to the total force), and
\begin{widetext}
\begin{alignat}{2}\label{eq:selfF}
\hspace{-20pt}
\fV^{\mbox{\tiny{source}}}[\qV,\vV;{\color{blue}\aV}](t)
  =  - \frac{\drm}{\drm{t}} \displaystyle\int_{B_{ct}(\qV_0)}
\Bigl( \PiV^{\mbox{\tiny{MBLTP}}}_{\mbox{\tiny{source}}}(t,\sV) - 
\PiV^{\mbox{\tiny{MBLTP}}}_{\mbox{\tiny{source}}}(0,\sV-\qV_0-\vV_{\!0}t)\Bigr) d^3{s}  
\end{alignat}
\end{widetext}
with $\PiV^{\mbox{\tiny{MBLTP}}}_{\mbox{\tiny{source}}}$
given by \refeq{eq:PiMBLTP} with $(\BV_1,\DV_1,\EV_1,\HV_1)$ in place of $(\BV,\DV,\EV,\HV)$.
\begin{rem}
 We re-emphasize that there is no such thing as \underline{\emph{the self-field force}} in electrodynamics; only the total force, i.e. 
the sum at r.h.s.\refeq{eq:totalF}, has an absolute meaning (in the chosen Lorentz frame).
 In particular, our ``self''-field force does generally not agree with the expression studied in \cite{Zayats} and \cite{GratusETal}, 
which depends on the complete \emph{actual} past motion of the point particle and cannot be used to study its initial
value problem in which only the particle's initial position and velocity are prescribed (given the initial fields).
\end{rem}

 The ``self''-field force can be evaluated using retarded spherical coordinates $(r,\vartheta,\varphi)$ to carry out the
$\drm^3{s}$ integrations over the ball ${B_{ct}(\qV_0)}$, after which one can differentiate w.r.t. $t$.
 This yields
\begin{widetext}
\begin{alignat}{2}\label{eq:selfFexpl}
\hspace{-20pt}
\fV^{\mbox{\tiny{source}}}[\qV,\vV;{\color{blue}\aV}](t)
&=  \frac{e^2} {4\pi } \biggl[ \biggr.
 - {\mathbf{Z}}_{\boldsymbol{\xi}}^{[2]}(t,t) + {\mathbf{Z}}_{\boldsymbol{\xi}^\circ}^{[2]}(t,t) \, 
\\ \notag
& \qquad\quad  -\!\!\! \;{\textstyle\sum\limits_{0\leq k\leq 1}}\! c^{2-k}(2-k)\!\!
\displaystyle  \int_0^{t}\! 
\Bigl[{\mathbf{Z}}_{\boldsymbol{\xi}}^{[k]}\big(t,t^{\mathrm{r}}\big)
-\!
{\mathbf{Z}}_{\boldsymbol{\xi}^\circ}^{[k]}\big(t, t^{\mathrm{r}}\big)\Bigr]
(t- t^{\mbox{\tiny{r}}})^{1-k} \drm{t^{\mbox{\tiny{r}}}} 
\\ \notag
& \qquad\quad -\!\!\!  \;{\textstyle\sum\limits_{0\leq k\leq 2}}\! c^{2-k}\!
\displaystyle  \int_0^{t}\! 
\Bigl[\tpddt{\mathbf{Z}}_{\boldsymbol{\xi}}^{[k]}\big(t,t^{\mathrm{r}}\big)
-\!
\tpddt{\mathbf{Z}}_{\boldsymbol{\xi}^\circ}^{[k]}\big(t, t^{\mathrm{r}}\big)\Bigr]
(t- t^{\mathrm{r}})^{2-k} \drm{t^{\mathrm{r}}}  \biggl. \biggr],
\end{alignat}
where $\boldsymbol{\xi}(t) \equiv (\qV,\vV,{\color{blue}\aV})(t)$ and 
$\boldsymbol{\xi}^\circ(t) \equiv (\qV_0+\vV_0t,\vV_0,{\boldsymbol{0}})$, where 
${\mathbf{Z}}_{\boldsymbol{\xi}}^{[2]}(t,t) :=\lim_{t^{\mathrm{r}}\to t}
{\mathbf{Z}}_{\boldsymbol{\xi}}^{[k]}\big(t,t^{\mathrm{r}}\big)$,
and where
\begin{alignat}{1}\label{Zdef}
{\mathbf{Z}}_{\boldsymbol{\xi}}^{[k]}\big(t,t^{\mathrm{r}}\big) = 
 \displaystyle  \int_0^{2\pi}\!\! \int_0^{\pi}\!
\left(1-\tfrac1c\abs{\vV^{}(t^\mathrm{r})}\cos\vartheta\right)
 \boldsymbol{\pi}_{\boldsymbol{\xi}}^{[k]}\big(t,\qV(t^\mathrm{r}) + c(t-t^\mathrm{r})\nV 
\big) 
\sin\vartheta \drm{\vartheta}\drm{\varphi}\,,
 \end{alignat}
with  
$\nV =\left(\sin\vartheta \cos\varphi,\;\sin\vartheta \sin\varphi ,\; \cos\vartheta \right)$
a normal vector to the retarded sphere of radius $r=c(t-t^\mathrm{r})$.
 Also, setting
\begin{alignat}{1}\label{rmK}
\mathrm{K}_{\boldsymbol{\xi}}(t',t,\sV) & :=
\tfrac{J_1\!\bigl(\varkappa\sqrt{c^2(t-t')^2-|\sV-\qV(t')|^2 }\bigr)}{\sqrt{c^2(t-t')^2-|\sV-\qV(t')|^2}^{\phantom{n}}},\\
\mathbf{K}_{\boldsymbol{\xi}}(t',t,\sV) & := 
\tfrac{J_2\!\bigl(\varkappa\sqrt{c^2(t-t')^2-|\sV-\qV(t')|^2 }\bigr)}{{c^2(t-t')^2-|\sV-\qV(t')|^2}^{\phantom{n}} }
 \left(\sV-\qV(t')- \vV(t')(t-t')\right), \label{bfK}
\end{alignat}
the $\boldsymbol{\pi}_{\boldsymbol{\xi}}^{[k]}(t,\sV)$ with $k\in\{0,1,2\}$ read, for $\sV\neq\qV$,
\begin{alignat}{1}
\label{pi0}
 \boldsymbol{\pi}_{\boldsymbol{\xi}}^{[0]}(t,\sV) =
& - \varkappa^4 \frac14\left[
{\textstyle{
\frac{\left({\nV(\qV,\sV)} -\frac1c{\vV}\right)\crprd\color{black}\left(\frac1c{\vV}\crprd {\nV(\qV,\sV)} \right)}{
      \bigl({1-\frac1c {\vV}\cdot\nV(\qV,\sV)}\bigr)^{\!2} }
           }}\right]_{\mathrm{ret}}\\ \notag
&+ \varkappa^4\frac12\left[
{\textstyle{
\frac{ {\nV(\qV,\sV)}
-\frac1c{\vV}}{ {1-\frac1c {\vV}\cdot\nV(\qV,\sV)} }
             }}\right]_{\mathrm{ret}} 
\crprd \!
\int_{-\infty}^{t^\mathrm{ret}_{\boldsymbol{\xi}}(t,\sV)}\!\!\!\!
{\vV(t')}\crprd \mathbf{K}_{\boldsymbol{\xi}}(t',t,\sV)\drm{t'} 
\\ \notag
& - \varkappa^4\frac12\left[{\textstyle{\frac{ \color{black} \frac1c{\vV}\crprd {\nV(\qV,\sV)} }{
      1-\frac1c {\vV}\cdot\nV(\qV,\sV)} }}\right]_{\mathrm{ret}} 
\crprd \int_{-\infty}^{t^\mathrm{ret}_{\boldsymbol{\xi}}(t,\sV)}\!\!\!\!
 c\mathbf{K}_{\boldsymbol{\xi}}(t',t,\sV)\drm{t'} 
\\ \notag
& - \varkappa^4 \int_{-\infty}^{t^\mathrm{ret}_{\boldsymbol{\xi}}(t,\sV)} \!\!\!\!
 c\mathbf{K}_{\boldsymbol{\xi}}(t',t,\sV)\drm{t'} \crprd \int_{-\infty}^{t^\mathrm{ret}_{\boldsymbol{\xi}}(t,\sV)} \!\!\!\!
{\vV(t')}\crprd \mathbf{K}_{\boldsymbol{\xi}}(t',t,\sV)\drm{t'} 
\\ \notag
& - \varkappa^4 c\int_{-\infty}^{t^\mathrm{ret}_{\boldsymbol{\xi}}(t,\sV)} \!\!\!\! \mathrm{K}_{\boldsymbol{\xi}}(t',t,\sV)\drm{t'} 
 \int_{-\infty}^{t^\mathrm{ret}_{\boldsymbol{\xi}}(t,\sV)}  \!\!\!\!\mathrm{K}_{\boldsymbol{\xi}}(t',t,\sV) {\vV}(t')\drm{t'}\,,
\end{alignat}
\begin{alignat}{1}
\label{pi1}
 \boldsymbol{\pi}_{\boldsymbol{\xi}}^{[1]}(t,\sV) = 
& - \varkappa^2 
\left[
{\textstyle{
{\color{black}
{\nV(\qV,\sV)}\frac{\left({\nV(\qV,\sV)}\crprd [{ \left({\nV(\qV,\sV)}-\frac1c{\vV}\right)\crprd {\color{blue}\aV} }]\right)\cdot\frac1c\vV}{
       c^2 \bigl({1-\frac1c {\vV}\cdot\nV(\qV,\sV)}\bigr)^{\!4} }
    } +
{\nV(\qV,\sV)} \crprd\frac{ \left({\nV(\qV,\sV)}-\frac1c{\vV}\right)\crprd {\color{blue}\aV} }{
      2c^2 \bigl({1-\frac1c {\vV}\cdot\nV(\qV,\sV)}\bigr)^{\!3} }
}}\right]_{\mathrm{ret}}\\ \notag
&- \varkappa^2\left[
{\textstyle{
{\nV(\qV,\sV)}\crprd\frac{ \left({\nV(\qV,\sV)}-\frac1c{\vV}\right)\crprd {\color{blue}\aV} }{
      c^2 \bigl({1-\frac1c {\vV}\cdot\nV(\qV,\sV)}\bigr)^{\!3} }
}}\right]_{\mathrm{ret}} \!\!
\crprd \!
\int_{-\infty}^{t^\mathrm{ret}_{\boldsymbol{\xi}}(t,\sV)}\!\!\!\!
{\vV(t')}\crprd \mathbf{K}_{\boldsymbol{\xi}}(t',t,\sV)\drm{t'} 
\\ \notag
& + \varkappa^2\left[\nV(\qV,\sV)\crprd \biggl[{\textstyle{\nV(\qV,\sV)\crprd 
\frac{\left({\nV(\qV,\sV)}_{\phantom{!\!}}-\frac1c{\vV}\right)\crprd{\color{blue}\aV} }{
      c^2\bigl({1-\frac1c {\vV}\cdot\nV(\qV,\sV)}\bigr)^{\!3} }
}}\biggr]\right]_{\mathrm{ret}} \!\!\! 
\crprd\! \int_{-\infty}^{t^\mathrm{ret}_{\boldsymbol{\xi}}(t,\sV)} \!\!\!\!
 c\mathbf{K}_{\boldsymbol{\xi}}(t',t,\sV)\drm{t'} 
\\ \notag
& {\color{red}+}\varkappa^3 
 \left[\textstyle\frac{1}{{1-\frac1c {\vV}\cdot\nV(\qV,\sV)} }\right]_{\mathrm{ret}} 
 \int_{-\infty}^{t^\mathrm{ret}_{\boldsymbol{\xi}}(t,\sV)}\!\!\!\!
 \mathrm{K}_{\boldsymbol{\xi}}(t',t,\sV)\left[{\vV}({t^\mathrm{ret}_{\boldsymbol{\xi}}(t,\sV)}))+{\vV}(t')\right]\drm{t'}\,,
 \end{alignat}
\begin{alignat}{1}
\label{pi2}
\boldsymbol{\pi}_{\boldsymbol{\xi}}^{[2]}(t,s) = & - \varkappa^2
\left[\textstyle\frac{1}{\bigl({1-\frac1c {\vV}\cdot\nV(\qV,\sV)}\bigr)^{\!2} }\frac1c{\vV}
{\color{black} - \Big[\!{1-\tfrac{1}{c^2}\big|\vV\big|^2}\!\Big]
\frac{ \left({\nV(\qV,\sV)}-\frac1c{\vV}\right) \crprd \left(\frac1c\vV\crprd \nV(\qV,\sV)\right) }{
      \bigl({1-\frac1c {\vV}\cdot\nV(\qV,\sV)}\bigr)^4 }}
\right]_{\mathrm{ret}} 
\\ \notag
&  +\varkappa^2 \left[\Big[\!{1-\tfrac{1}{c^2}\big|\vV\big|^2}\!\Big]\nV(\qV,\sV)\crprd {\textstyle{
\frac{ {\nV(\qV,\sV)}_{\phantom{!\!}}-\frac1c{\vV} }{
      \bigl({1-\frac1c {\vV}\cdot\nV(\qV,\sV)}\bigr)^{\!3} }
}}\right]_{\mathrm{ret}} \crprd\int_{-\infty}^{t^\mathrm{ret}_{\boldsymbol{\xi}}(t,\sV)} \!\!\!\!
 c\mathbf{K}_{\boldsymbol{\xi}}(t',t,\sV)\drm{t'} \\
\notag
& -  \varkappa^2\left[\Big[\!{1-\tfrac{1}{c^2}\big|\vV\big|^2}\!\Big]
{\textstyle{
\frac{ {\nV(\qV,\sV)}_{\phantom{!\!}}-\frac1c{\vV} }{
      \bigl({1-\frac1c {\vV}\cdot\nV(\qV,\sV)}\bigr)^{\!3} }
}}\right]_{\mathrm{ret}} 
\crprd \!
\int_{-\infty}^{t^\mathrm{ret}_{\boldsymbol{\xi}}(t,\sV)} \!\!\!\!
{\vV(t')}\crprd \mathbf{K}_{\boldsymbol{\xi}}(t',t,\sV)\drm{t'} ,
\end{alignat}
and $\big|_\mathrm{ret}$ means that $\qV(\tilde{t})$, $\vV(\tilde{t})$, ${\color{blue}\aV}(\tilde{t})$ 
are evaluated at~$\tilde{t} = {t^\mathrm{ret}_{\boldsymbol{\xi}}}(t,\sV)$.
 Note that r.h.s.\refeq{eq:selfFexpl} vanishes at $t=0$.
\end{widetext}

 Several of the spherical angular integrations can be carried out explicitly in terms of
well-known functions; see \cite{KTZonBLTP}. 
 Here we are content with the remark that for $C^{1,1}$ motions $t\mapsto\qV(t)$ (which for $t\leq 0$ 
coincide with the auxiliary straight line motion $t\mapsto \qV(0) + \vV(0)t$ employed to \emph{define} the field initial data,
and should not be confused with any actual motion for $t<0$),
one can easily show that all the terms in $\fV^{\mbox{\tiny{source}}}[\qV,\vV;{\color{blue}\aV}](t)$ are well-defined. 
 Since \refeq{eq:vacuumF} is also well-defined for the regular vacuum field, the electrodynamical
admissibility of the BLTP vacuum law follows.

	\section{BLTP electrodynamics has no $\dddot\qV$-problem}\label{sec:BLTP}\vspace{-0.2truecm}
%
 Having established that the BLTP law is electrodynamically admissible, in the sense that the solutions
of the MBLTP field equations for prescribed $C^{1,1}$ motions $t\mapsto\qV(t)$ 
yield a field momentum vector density which is integrable over space, and its integral differentiable in time,
we now explain that the so obtained electromagnetic force \refeq{eq:totalF}, with 
$\fV^{\mbox{\tiny{vacuum}}}[\qV,\vV](t)$ given by \refeq{eq:vacuumF} and
$\fV^{\mbox{\tiny{source}}}[\qV,\vV;{\color{blue}\aV}](t)$ by \refeq{eq:selfFexpl}ff, 
when substituted at r.h.s.\refeq{eq:NewtonSECONDlaw}, yields a well-posed initial value problem for point charge motion, 
and thus a well-posed classical BLTP electrodynamics with a point charge. 
 It is understood here that the relativistic velocity-momentum relation \refeq{eq:MOMinTERMSofVELO} is assumed,
 with $\mbare\neq 0$.
 This well-posedness result is a special case of a result
proved in \cite{KTZonBLTP} for an arbitrary finite number of point charges.
 We here explain the main idea of the proof for fields having a single point charge source. 

 A key feature of the total electrodynamical force in a BLTP vacuum, 
$\fV^{\mbox{\tiny{vacuum}}}[\qV,\vV] + \fV^{\mbox{\tiny{source}}}[\qV,\vV;{\color{blue}\aV}]$, as indicated by our notation,
is the dependence on only $\qV,\vV,{\color{blue}\aV}$; a third- (and higher-)order time derivative of $\qV(t)$ does not show up. 
 Therefore, already purely formally, our initial value problem for the point charge motion is of second order, as desired.

 Next, inserting \refeq{eq:MOMinTERMSofVELO} at l.h.s.\refeq{eq:NewtonSECONDlaw} and carrying out the differentiation 
we obtain a familiar expression which can be rewritten as a regular matrix acting on the vector ${\color{blue}\aV}$,
the matrix depending only on $\pV$ (and $\mbare$), not on $\qV$ and not on ${\color{blue}\aV}$. 
 Applying the inverse of this matrix at both sides of \refeq{eq:NewtonSECONDlaw} (with the force in a BLTP vacuum in place),
our equation of motion becomes
\begin{equation}\label{eq:Volterra}
{\color{blue}\aV}
= 
 W[\pV]\cdot \Big( 
\fV^{\mbox{\tiny{vacuum}}}[\qV,\vV]
+
\fV^{\mbox{\tiny{source}}}[\qV,\vV;{\color{blue}\aV}]\Big)
\end{equation}
where the velocity $\vV$ is expressed in terms of the momentum $\pV$ by inverting \refeq{eq:MOMinTERMSofVELO}, i.e.
\begin{equation}\label{ELPinverse}
\vV = \frac{1}{\mbare} \frac{\pV}{\sqrt{1 +\frac{|\pV|^2}{\mbare^2 c^2 }}};\quad \mbare\neq 0,
\end{equation}
and  where  
\begin{equation}
W[\pV]
:= 
\frac{1}{\mbare} \frac{1}{\sqrt{1 +\frac{|\pV|^2}{\mbare^2 c^2 }}} 
\left[\ONE - \frac{\pV\otimes \pV}{{{\mbare^{2}c^2} + {|\pV|^2}}}\right].
\end{equation}
 Together with the usual definitions ${\color{blue}\aV}(t):=\Ddt\vV(t)$ and $\vV(t):=\Ddt\qV(t)$ this is a complicated, nonlinear 
and implicit, second-order differential-integral equation of motion for the position of the point charge.

 We approach the problem stepwise. 
 Temporarily ignoring the relationships ${\color{blue}\aV}(t)=\Ddt\vV(t)$ and $\vV(t)=\Ddt\qV(t)$, and instead treating the
maps $t\mapsto\qV(t)$ and $t\mapsto\pV(t)$ and $t\mapsto {\color{blue}\aV}(t)$ as a-priori unrelated for $t>0$ (of course,
for $t\leq 0$ the maps are determined by the stipulated auxiliary motion), we note that then
the acceleration $t\mapsto {\color{blue}\aV}(t)$ enters the BLTP ``self'' force term 
$\fV^{\mbox{\tiny{source}}}[\qV,\vV;{\color{blue}\aV}]$ only in a \emph{linear} fashion. 
 Therefore, when treating $t\mapsto\qV(t)$ and $t\mapsto\vV(t)$ as given, \refeq{eq:Volterra} becomes 
a linear integral equation for the acceleration $t\mapsto {\color{blue}\aV}(t)$. 
 Better yet, inspection shows that it is a \emph{linear Volterra integral equation} for which we 
prove the following key result, see \cite{KTZonBLTP}.

\noindent
\begin{prp} Given $C^{0,1}$ maps $t\mapsto{\qV}(t)$ and $t\mapsto{\pV}(t)$, with Lip$(\qV)=v$, Lip$(\vV)=a$, and
$|\vV(t)|\leq v<c$, which for $t\leq 0$ coincide with the stipulated unaccelerated auxiliary motion,
the Volterra equation as a fixed point map has a unique $C^0$ solution 
$t\mapsto {\color{blue}\aV}(t) = \boldsymbol{\alpha}[\qV(\,\cdot\,),\pV(\,\cdot\,)](t)$. 
 Moreover, the solution depends Lip\-schitz continuously on the maps
$t\mapsto{\qV}(t)$ and $t\mapsto{\pV}(t)$. 
\end{prp}
 The careful proof in \cite{KTZonBLTP} fills many pages.

 Essentially as a corollary of the above Proposition we obtain the well-posedness result for the joint initial value problem 
of the MBLTP field and its point charge source.
 Indeed, now substituting $\boldsymbol{\alpha}[\qV(\,\cdot\,),\pV(\,\cdot\,)](t)$ for ${\color{blue}\aV}(t)$ in 
$\fV^{\mbox{\tiny{source}}}[\qV,\vV;{\color{blue}\aV}]$, Newton's equation of motion, supplemented with the 
relativistic velocity-momentum relation \refeq{ELPinverse} and the definition $\vV(t)=\Ddt\qV(t)$ of the velocity,
can be formally integrated to become a fixed point map for
a curve in the phase space of the point charge, viz. for some $T>0$ and $t\in[0,T)$,
\begin{alignat}{1}
\hspace{-20pt}
\qV(t)  = \label{eq:iteraQ}
\qV(0) & + \int_0^t \frac{1}{\mbare^{}} \frac{\, \pV({t'})}{\sqrt{1+ \frac{|\pV({t'})|^2}{\mbare^{2}c^2}}}\drm{{t'}},\\
\hspace{-25pt}
\pV(t)= \label{eq:iteraP}
\pV(0) & + \int_0^t\!  \fV^{\mbox{\tiny{vacuum}}}[\qV,\vV]\!(t')\,\drm{t'} \\ \notag
 & + \int_0^t\! \fV^{\mbox{\tiny{source}}}\big[\qV,\vV;\boldsymbol{\alpha}[\qV(\,\cdot\,),\pV(\,\cdot\,)]\big]\!(t')\,\drm{t'},
\hspace{-5pt}
\end{alignat}
where $\vV$ at r.h.s.\refeq{eq:iteraP} is given in terms of $\pV$ through \refeq{ELPinverse}.
 The following theorem is a special case of the $N$-body result  proved in \cite{KTZonBLTP}.
\begin{thm}
 Given $\qV(0)$ and $\pV(0)$ and the stipulated MBLTP field initial data, there is a $T>0$ such that 
the fixed point problem \refeq{eq:iteraQ} \&\ \refeq{eq:iteraP} has a unique $C^{1,1}$ solution $t\mapsto(\qV,\pV)(t)$
for $t\in(0,T)$ extending continuously to $[0,T)$.
 Moreover, if in a finite time the particle does not reach the speed of light or infinite acceleration, then $T=\infty$.
 In any event, total energy-momentum conservation holds.
\end{thm}
\begin{rem}
 When $\fV^{\mbox{\tiny{vacuum}}}[\qV,\vV]$ is replaced by an external, smooth and short-ranged force field,
the dynamics is global, i.e. $T=\infty$. 
 This was shown in \cite{VuMaria}.
\end{rem}
\begin{rem}
 For the BLTP electrodynamics it is easy to generalize the single-particle formulation to the $N$-body formulation.
 By the linearity of the MBLTP field equations we can associate each particle with its own Li\'enard--Wiechert(-type) 
 MBLTP field, and thus a ``self''-field force, same expressions as before except that the index ${}_1$ is replaced by ${}_j$, 
say, also to be attached to position, velocity, momentum, and acceleration vectors, and to the charge and mass parameters.
 The total MBLTP field is the sum of a vacuum field of the kind considered above plus all the Li\'enard--Wiechert(-type) 
 fields.
 The total force now gets an extra contribution in form of the Lorentz force \refeq{eq:Lf} on particle ${}_j$ exerted 
by the Li\'enard--Wiechert-type fields of all the particles but ${}_j$.
 This requires extra regularity estimates for all $\EV_n$ and $\BV_n$, $n\in\{1,...,N\}$, 
which are established in \cite{KTZonBLTP}.
 In addition to the avoidance of the possible finite-time blow-up scenarios for the single particle dynamics, 
global well-posedness now also requires that in finite time  no two particles reach the same location.
 While this can conceivably happen, it is not known whether it will happen generically or not.
\end{rem}

\begin{rem}
  Above we noted already that our ``self''-field force r.h.s.\refeq{eq:selfFexpl} vanishes at the initial instant $t=0$.
  Thus in a BLTP vacuum the inertia of the particle is initially entirely due to its bare mass! 
\end{rem}

 To summarize, the work of \cite{KTZonBLTP} on BLTP electrodynamics demonstrates that it is feasible to set up a
classical electrodynamics with point charge sources as a well-posed joint initial value problem for the fields and the
particles, which is of second order in the particle positions.
 No Landau--Lifshitz type approximation has been invoked because the infamous $\dddot\qV$-problem does not show up;
no negative infinite bare mass renormalization \cite{DiracA}, no additional comparison axioms \cite{QuinnWald} (cf. also
\cite{Kijowski}), and no ``separating off of singularities'' \cite{DetweilerWhiting} are invoked. 

 Whether BLTP electrodynamics is already a physically acceptable classical theory is a different question.
 Particularly embarrassing is the fact that the triumph of avoiding unphysical third- (and higher-)order time derivatives
of the point charge's position $\qV(t)$ in its force law is paid for by a high prize, namely by introducing (presumably) unphysical 
higher-order time derivatives in the electromagnetic field equations! 
 As a consequence, four {MBLTP} {fields}, namely $\BV,\DV,\EV,\dot\EV$, {require initial data}.
 However, according to phenomenological electromagnetism, once the fields 
$\BV(0,\sV)$ and $\DV(0,\sV)$ are determined / prescribed initially (constrained by (\ref{eq:MdivB}) and (\ref{eq:MdivD})),\footnote{In 
   the physics literature on classical Lorentz electrodynamics, one usually finds $\BV(0,\sV)$ and $\EV(0,\sV)$ prescribed, but recall that 
   $\DV=\EV$ (and $\BV=\HV$) in Lorentz electrodynamics.} 
one does not have any freedom left to also choose $\EV_0(0,\sV)$ and $\dot\EV_0(0,\sV)$, yet
the MBLTP field equations do require such a choice.
 (Incidentally, neither the founding fathers of the BLTP theory, nor Feynman \cite{FeynmanBOPP}, nor recent authors
\cite{Zayats}, \cite{GratusETal}, seem to have been worried about these ``loose ends'' of this field theory.)
 Be that as it may, it is generally agreed upon, and mathematically realized in the structures of
the Maxwell--Lorentz field equations and of the Maxwell--Born--Infeld field equations,
that once the fields $\BV(0,\sV)$ and $\DV(0,\sV)$ are prescribed, the initial field data are fixed. 
 Therefore, to implement this rule also into the MBLTP field theory one needs a prescription which expresses 
the data $\EV(0,\sV)$ and $\dot\EV(0,\sV)$ in terms of $\BV(0,\sV)$ and $\DV(0,\sV)$. 

 In \cite{KTZonBLTP} we show that this can be accomplished by
postulating that the fields $\EV_0(0,\sV)$ and $\dot\EV_0(0,\sV)$ maximize the field energy functional initially, given 
$\BV_0(0,\sV)$, $\DV_0(0,\sV)$, and given the Li\'enard--Wiechert(-type) fields associated with the data $\qV(0)$, $\vV(0)$.
 Thus the initial field energy is made as little negative as possible.
 One can make this constraint Lorentz invariant 
by stipulating, e.g., that the field energy maximization refers to the Lorentz frame in which the total particle momentum vanishes.
 In the Lorentz frame in which the initial value problem is formulated the relevant constraint is obtained through
a Lorentz boost.

	\section{The electromagnetic force: \hspace{1truecm} $N$ point charges}\label{sec:EMfN}\vspace{-5pt}
%
 In the case of BLTP electrodynamics it is straightforward to generalize the formula for the electromagnetic
force on a point charge from when there is only a single charge to when in total $N$ point charges are present.
 This is possible because  the linear pre-metric Maxwell equations are then complemented with 
the linear BLTP vacuum law as closure relation.
 When nonlinear vacuum laws are used, for instance the Born--Infeld law, all these linear algebra-based conclusions
are not available. 
 In the following we first give a general \emph{distributional} definition of the electromagnetic force on a point 
charge source when $N$ point charges are present.
 Then we extract from this definition the $N$-body analog of the
earlier given basic definition \refeq{eq:EMf} of the electromagnetic force when only a single charge is present. 

	\subsection{Distributional definition of the electromagnetic force}\label{sec:fDISTRIB}\vspace{-5pt}
 In the electromagnetic field-theory part of electrodynamics one already implements the fact that 
the system of $N$ moving charged point particles is associated with a distribution-valued four-vector field on spacetime,
namely $(c\rho,\jV)(t,\sV)$, the inhomogeneity term in the linear pre-metric Maxwell field equations.
 It is only natural to also write the mechanical quantities associated with the point particles as 
distribution-valued fields on spacetime and thereby treat particles and fields on an equal footing.

 The \emph{momentum vector-density}
\begin{equation}
\PiV^{\mbox{\tiny{charge}}}
(t,\sV)
= \label{eq:TokCHARGES}
{\sum_n}\, \pV_{n}(t)\; \delta_{\qV_n(t)}(\sV)
\end{equation}
and the \emph{symmetric stress tensor field}
\begin{equation}
T^{\mbox{\tiny{charge}}} (t,\sV)
= \label{eq:TchargeSTRESS}
\sum_n  
\frac{1}{m_n^{}}\frac{\pV_{n}(t)\otimes \pV_{n}(t)}{\sqrt{1 + \frac{|\pV_{n}(t)|^2}{m_n^{2}c^2}}}\; \delta_{\qV_n(t)}(\sV)
\end{equation}
jointly satisfy the \emph{local law of particle momentum balance} 
\begin{equation}
\textstyle
\pddt \PiV^{\mbox{\tiny{charge}}} (t,\sV) + \nabla\cdot T^{\mbox{\tiny{charge}}} (t,\sV)
= \label{eq:chargeMOMbalanceLAW}
 \sum_{n}  \fV^{\mbox{\tiny{em}}}_n(t)\delta_{\qV_n(t)}(\sV);
\end{equation}
here, $\fV^{\mbox{\tiny{em}}}_n(t)$ is precisely the force term in \refeq{eq:NewtonSECONDlaw}, now understood for the $n$-th particle.

 Similarly, for admissible vacuum laws, the \emph{electromagnetic field momentum vector-density} $\PiV^{\mbox{\tiny{field}}}$
and the \emph{symmetric stress tensor of the fields}\footnote{To avoid awkward minus signs elsewhere, we 
      define $T^{\mbox{\tiny{field}}}$ with the opposite sign compared to the 
      convention introduced by Maxwell in what nowadays is called the ``Maxwell stress tensor.''}
$T^{\mbox{\tiny{field}}}$ 
jointly satisfy the \emph{local law of field momentum balance} 
\begin{equation}
\textstyle
\pddt \PiV^{\mbox{\tiny{field}}} (t,\sV) + \nabla\cdot T^{\mbox{\tiny{field}}} (t,\sV)
= \label{eq:fieldMOMbalanceLAW}
 \textstyle\sum\limits_{n}  \gV_n(t)\delta_{\qV_n(t)}(\sV).
\end{equation}
 The source (sink) terms, $\gV_n(t)$,  for the electromagnetic field momentum density \&\ field stress
are \emph{defined} by \refeq{eq:fieldMOMbalanceLAW}.
 For admissible vacuum laws \refeq{eq:fieldMOMbalanceLAW} is indeed \emph{well-defined in the sense of distributions} 
for fields with point charge sources at the $\qV_n(t)$ which move with subluminal velocities $\vV_{n}(t)$.
 The vector $\gV_n(t)$ is the electromagnetic field momentum gained (lost) per unit of time at 
location $\qV_n^{}(t)$ of  the source (sink).
 Therefore $\gV_n(t)$ has the physical dimension of a ``force.''

 Now we \emph{postulate} that the total momentum vector-density of the interacting system of electromagnetic field 
and its charged particle sources is given by the sum of the respective field and particle expressions defined earlier,
\begin{equation}\label{eq:totalMOMENTUMdef}
\PiV(t,\sV):=\PiV^{\mbox{\tiny{field}}}(t,\sV) + \PiV^{\mbox{\tiny{charge}}}(t,\sV)
\end{equation}
is the total momentum vector-density.
 Similarly, the symmetric total stress tensor is \emph{postulated} to be the sum
\begin{equation}\label{eq:totalSTRESSdef}
T(t,\sV):=T^{\mbox{\tiny{field}}}(t,\sV)+T^{\mbox{\tiny{charge}}}(t,\sV).
\end{equation}
 Our postulates \refeq{eq:totalMOMENTUMdef} \&\ \refeq{eq:totalSTRESSdef} in concert with the two balance laws 
\refeq{eq:chargeMOMbalanceLAW} \&\ \refeq{eq:fieldMOMbalanceLAW} imply the \emph{local balance law for the total momentum
vector-density},
\begin{equation}
\textstyle
\pddt \PiV(t,\sV) + \nabla\cdot T (t,\sV)
= \label{eq:MOMtotalBALANCE}
 \textstyle\sum\limits_{n}  \big(\gV_n(t) + \fV^{\mbox{\tiny{em}}}_n(t)\big) \delta_{\qV_n(t)}(\sV).
\end{equation}

 Now \emph{postulating} that the total momentum density and the stresses jointly satisfy the
\emph{local conservation law for the total momentum vector-density}, 
\begin{equation}
\textstyle
\pddt \PiV(t,\sV) + \nabla\cdot T (t,\sV)
= \label{eq:MOMconservation}
\NullV,
\end{equation}
then by comparing \refeq{eq:MOMconservation} with \refeq{eq:MOMtotalBALANCE}, and invoking relativistic locality
(spacelike separated events do not affect each other), one \emph{deduces} the identities
\begin{equation}\label{eq:EMfN}
\forall n\, \&\ \forall t \geq 0\, (a.e.):\ \fV^{\mbox{\tiny{em}}}_n(t) \equiv - \gV_n(t).
\end{equation}
 This is the general distributional definition of the electromagnetic force on a point charge source of the classical
electromagnetic field when $N$ point charges are present.
 It may be seen as a zero-gravity implementation of what Einstein--Infeld--Hoffmann surmised \cite{EIH}.

 In the next subsection we will extract the $N$-body analog of \refeq{eq:EMf}, the electromagnetic force when only a single 
charge is present. 

 We close this subsection by recalling that the \emph{energy density of the charged particle distribution} 
\begin{equation}
 \veps^{\mbox{\tiny{charge}}} (t,\sV)
= \label{eq:TooCHARGES}
 \sum_n m_n^{}c^2 {\sqrt{1 + \tfrac{|\PV_{\!\!n}(t)|^2}{m_n^{2}c^2}}}\; \delta_{\QV_n(t)}(\sV)
\end{equation}
and its momentum vector-density \refeq{eq:TokCHARGES} jointly satisfy the \emph{local law of particle energy balance}
\begin{equation}
\textstyle
\pddt \veps^{\mbox{\tiny{charge}}} (t,\sV) + c^2 \nabla\cdot \PiV^{\mbox{\tiny{charge}}} (t,\sV) 
= \label{eq:chargeENbalanceLAW}
 \textstyle\sum\limits_{n}  \fV^{\mbox{\tiny{em}}}_n(t)\cdot {\vV_{n}(t)} \delta_{\qV_n(t)}(\sV).
\end{equation}
 Similarly,  the \emph{field energy density} $\veps^{\mbox{\tiny{field}}}(t,\sV)$
and the field momentum density $\PiV^{\mbox{\tiny{field}}} (t,\sV)$ jointly satisfy the \emph{local law of field energy balance}
\begin{equation}
\textstyle
\pddt \veps^{\mbox{\tiny{field}}} (t,\sV) + c^2 \nabla\cdot \PiV^{\mbox{\tiny{field}}} (t,\sV) 
= \label{eq:fieldENbalanceLAW}
\textstyle\sum\limits_{n}  \gV_n(t)\cdot {\vV_{n}(t)} \delta_{\qV_n(t)}(\sV).
\end{equation}
 When the total energy density is \emph{postulated} to be additive, i.e.
\begin{equation}\label{eq:totalENERGYdef}
\veps(t,\sV):= \veps^{\mbox{\tiny{field}}}(t,\sV) + \veps^{\mbox{\tiny{charge}}}(t,\sV),
\end{equation}
then the postulated local law of total momentum conservation \refeq{eq:MOMconservation} now entails 
the \emph{local conservation law for the total energy},
\begin{equation}
\textstyle
\pddt \veps (t,\sV) + c^2 \nabla\cdot \PiV(t,\sV) 
= \label{eq:ENconservation}
0.
\end{equation}

 Similarly one can obtain the \emph{local conservation law for total angular-momentum}.

	\subsection{Integral formula for the electromagnetic force}\label{sec:fDISTRIBeval}\vspace{-5pt}
 The distributional definition \refeq{eq:EMfN} of the electromagnetic force, with $\gV_n$ defined by \refeq{eq:fieldMOMbalanceLAW}, 
gives rise to the following $N$-body analog of the one-body formula \refeq{eq:EMf}. 
 Let $\cV_j$ denote the \emph{Voronoi cell} of the $j$-th point charge at the initial time, and $\partial\cV_j$ its boundary. 
 Let $T>0$ denote the instant of time until which all $N$ point charges remain in their initial Voronoi cells.
 Then integrating  \refeq{eq:fieldMOMbalanceLAW} over $\cV_j$ yields, for $t\in(0,T)$ and $\forall\,j\in\{1,...,N\}$, 
\begin{equation}
 \fV^{\mbox{\tiny{em}}}_j(t)
= \label{eq:EMfNint}
- \Ddt \int_{\cV_j} \PiV^{\mbox{\tiny{field}}} (t,\sV) \drm^3s 
- \int_{\partial\cV_j} (T^{\mbox{\tiny{field}}}\cdot\nu_j) (t,\sV) \drm^2s, 
\end{equation}
and the initial force is defined as its limit when $t\downarrow 0$.
 If $N=1$, so that $j=1$, we have $\cV_1 = \Rset^3$, the surface integral vanishes, and
in this case \refeq{eq:EMfNint} reduces to \refeq{eq:EMf}.

 If $T<\infty$ then to go beyond $T$ one can reset the clock to a new ``initial'' time, say $T-\eps$, 
and replace the initial Voronoi cells with those at time $t=T-\eps$, and repeat.
 
 As shown explicitly for BLTP electrodynamics, so also for BI electrodynamics one should be able to show that with fixed
initial data the r.h.s.\refeq{eq:EMfNint} depends on the vector functions $t'\mapsto\qV_n(t')$, $t'\mapsto\vV_n(t')$, 
and  $t'\mapsto {\color{blue}\aV}_n(t')$ for $t'\in(0,t)$, with $\lim_{t'\downarrow 0}\qV_n(t') = \qV_n(0)$
and $\lim_{t'\downarrow 0}\vV_n(t') = \vV_n(0)$ given particle initial data, i.e.
no higher-order time derivative of the position beyond the second one should show up.
 Then with $t'\mapsto\qV_n(t')$ and $t'\mapsto\vV_n(t')$ considered given, the system of relativistic Newton equations of motion
\refeq{eq:NewtonSECONDlaw} with \refeq{eq:EMfNint} at its right-hand side
becomes a system of generally nonlinear integral equations for the maps $t'\mapsto {\color{blue}\aV}_n(t')$ as functionals of 
the $\qV_n(\,.\,)$ and $\vV_n(\,.\,)$ (or $\pV_n(\,.\,)$). 
 Whenever this system has a unique solution which depends Lipschitz continuously on the $\qV_n(\,.\,)$ and $\vV_n(\,.\,)$ (or $\pV_n(\,.\,)$),
the electrodynamical initial value problem is locally (in time) well-posed. 

	\section{Motion along a constant electric field}\label{sec:EXAMPLES}

 The problem of determining the classical dynamics of a single point charge which moves along a static,
spatially homogeneous electric field (approximately achieved by the field between the plates 
of a charged capacitor) has already been mentioned in the introductory section, where we recalled that 
the Eliezer--Ford--O'Connell equation of motion, and also its Landau--Lifshitz approximation, 
fail to account for the radiation-reaction on the motion and merely reproduce the test particle motion.
 We now demonstrate that the initial-value problem for BLTP electrodynamics formulated in this paper 
does take the radiation-reaction on the motion into account.

 For simplicity we restrict the discussion to the case where the particle is initially at rest and
surrounded by its own electrostatic field and by the electrostatic field of the capacitor. 
 Note that this completely fixes the initial data for the field and for the particle.
 No ``past hypothesis,'' about how these initial data were established, is needed.
 
 The textbook idealization of a single point charge placed in a 
truly uniform capacitor field is obtained as a limiting case from our setup, as follows.
 Consider a system of $2N+1$ charges, with $j=1$ for the point electron whose dynamics we are interested in,
and $N$ positive and $N$ negative singly charged particles distributed uniformly over the respectively charged two
capacitor plates. 
 We  take the formal $N\to\infty$ limit in which the capacitor plates become
infinitely charged, but also infinitely extended and separated, leaving a homogeneous, 
static vacuum field $\EV^{\mathrm{hom}}$ behind in which the point charge ${}_1$ is situated. 
 While the total electrical field energy diverges in this limit, the field momentum is initially zero and 
remains well-defined later on; also the particle momentum vanishes initially for our data.
 The electromagnetic force on particle ${}_1$ is derived from total momentum \emph{balance} --- note that the total momentum 
of the single-particle-plus-field system is not by itself conserved because this is only a \emph{subsystem} of a formally 
``infinitely-many-particles-plus-field'' system.
 Even though all the other charges have been ``moved to spatial infinity,'' they still exert an influence on the remaining point
charge ${}_1$ through their field $\EV^{\mathrm{hom}}$.
 The balance equation for the momentum of the single-particle-plus-field subsystem thus reads 
\begin{equation}
\Ddt\left( \pV_1(t) + \pV^{\mbox{\tiny{field}}} (t)\right)
 = \label{eq:TOTALpBALANCEcapacitor}
-e \EV^{\mathrm{hom}} ,
\end{equation}
which yields the total electromagnetic force on particle~${}_1$,
\begin{equation}
 \fV^{\mbox{\tiny{em}}}_1(t)
= \label{eq:EMfCAPACITOR}
- \Ddt \int_{B_{ct}(\qV_0)} \PiV^{\mbox{\tiny{MBLTP}}} (t,\sV) \drm^3s 
-e \EV^{\mathrm{hom}} ,
\end{equation}
for $t\geq 0$.
 Technically we obtain \refeq{eq:EMfCAPACITOR} from \refeq{eq:EMfNint} for $j=1$.
 The point charge's Voronoi cell $\cV_1\to\Rset^3$ in the limit $N\to\infty$, and \refeq{eq:EMfNint} for $j=1$ gives \refeq{eq:EMfCAPACITOR}.
 We remark that the $\EV^{\mathrm{hom}}$ term in  \refeq{eq:EMfCAPACITOR} comes from the boundary integral in \refeq{eq:EMfNint};
it is easy to see that $\EV^{\mathrm{hom}}$ does not contribute to $\Ddt\int_{\Rset^3} \PiV^{\mbox{\tiny{MBLTP}}} (t,\sV) \drm^3s$.
 Moreover, in \refeq{eq:EMfCAPACITOR} we have replaced the domain of integration $\Rset^3$ by the ball $B_{ct}(\qV_0)$ because 
$\PiV^{\mbox{\tiny{MBLTP}}} (t,\sV) \equiv \NullV$ outside of this ball.
 The ``self''-field force $-\Ddt\int_{B_{ct}(\qV_0)} \PiV^{\mbox{\tiny{MBLTP}}} (t,\sV) \drm^3s$ is in this example identical 
with r.h.s.\refeq{eq:selfF}, for $\PiV^{\mbox{\tiny{MBLTP}}} (0,\sV)~\equiv~\NullV$.

 To evaluate $-\Ddt\int_{B_{ct}(\qV_0)} \PiV^{\mbox{\tiny{MBLTP}}} (t,\sV) \drm^3s$ 
we invoke equations \refeq{eq:selfFexpl}ff with ${\mathbf{Z}}_{\boldsymbol{\xi}^\circ}^{[2]}(t,t^{\mathrm{r}}) \equiv\NullV$.
 Due to the highly symmetrical setup these expressions simplify drastically, although one still cannot carry out each and 
every integration in terms of elementary functions --- but this is not necessary for our purposes here.

 We now consider first the early time regime.
 Since, as noted above, the ``self''-field force r.h.s.\refeq{eq:selfFexpl} vanishes at the initial instant $t=0$, 
the initial force is identical to the force $-e\,\EV^{\mathrm{hom}}$ on a charged test particle.
 Since the ``self''-field force varies continuously with $t\geq 0$, it will remain small for a certain amount of time 
(which we are not going to determine here). 
 During this early dynamical phase the motion of the particle is therefore well approximated by the test particle
dynamics, with the inertia of the particle given by its bare mass. 
 
 We now show that although the radiation-reaction force vanishes initially, it typically does not remain zero, unlike the
Eliezer--Ford--O'Connell radiation-reaction force and its Landau--Lifshitz approximation.
 By ``typically'' we mean that for almost all $\varkappa$ parameter values the radiation-reaction force does not vanish
(we do not rule out that there might be special, isolated values of $\varkappa$ for which this might happen). 
 For this purpose, assume to the contrary that the radiation-reaction force r.h.s.\refeq{eq:selfFexpl} would vanish for an open interval 
of $\varkappa$ values.
 Since r.h.s.\refeq{eq:selfFexpl}ff reveals that the radiation-reaction force is an analytic function of $\varkappa$, it 
follows that it then has to vanish for all $\varkappa$. 
 This in turn means that each and every Maclaurin coefficient of its power series expansion in $\varkappa$ has to vanish.
 {\color{orange}\cancel{But}} {\color{magenta}T}he lowest-order term, which is $\propto\varkappa^2$ and can be evaluated exactly in closed form (see below),
does {\color{orange}\cancel{not}} vanish, {\color{magenta}but the term $\propto\varkappa^3$ does not},
 hence the radiation-reaction force cannot typically vanish. 
 
 To compute the $O(\varkappa^2)$ contribution, divide the expressions for $\boldsymbol{\pi}^{[k]}_{\boldsymbol{\xi}}$ by $\varkappa^2$ and
take the limit $\varkappa\to 0$. 
 The only two terms which survive in the limit are those in the first line of r.h.s.\refeq{pi1} and r.h.s.\refeq{pi2},
respectively. 
 Carrying out the pertinent integrations in \refeq{Zdef}, and noting that the result only depends on $t^{\mathrm{r}}$, not on $t$,
so that the third line at r.h.s.\refeq{eq:selfFexpl}ff vanishes at $O(\varkappa^2)$, we obtain
\begin{widetext} 
\begin{alignat}{1}
\label{eq:SELFforceEXPLICITexpand}
 \fV^{\mbox{\tiny{em}}}_1(t) = 
& 
- \frac12 e^2 \varkappa^2  \frac{\vV(t)}{\abs{\vV(t)}} 
\left[ 2 \frac{c}{|\vV(t)|} - \frac{c^2}{|\vV(t)|^2} \ln \frac{1+\frac1c\abs{\vV(t)}}{1-\frac1c\abs{\vV(t)}} \right]
\\ \notag 
& - {\color{orange}\cancel{\frac32}}  e^2 \varkappa^2   \int_0^{t}\! 
\aV(t^{\mbox{\tiny{r}}})\frac{1}{|\vV(t^{\mbox{\tiny{r}}})|^2} \Biggl[ 
\frac{2 {\color{orange}\cancel{-\tfrac13\tfrac1c{|\vV(t^{\mbox{\tiny{r}}})|}}}  -  \frac{|\vV(t^{\mbox{\tiny{r}}})|^2}{c^2}}
{1-\frac{|\vV(t^{\mbox{\tiny{r}}})|^2}{c^2}}  - 
 \left(\frac{c}{|\vV(t^{\mbox{\tiny{r}}})|} {\color{orange}\cancel{-\frac16}} \right)
\ln\frac{1+\frac1c|\vV(t^{\mbox{\tiny{r}}})|}{1-\frac1c|\vV(t^{\mbox{\tiny{r}}})|}
\Biggr] \biggr.  c \drm{t^{\mbox{\tiny{r}}}}  + O(\varkappa^3) \\
\label{eq:SELFforceEXPLICITexpandMERGE}
= & {\color{orange}-\cancel{\frac12 e^2 \varkappa^2 \frac{\vV(t)}{\abs{\vV(t)}} }
\left[\cancel{1-\frac{c}{|\vV(t)|}\left(1+\left[1-\frac{c}{|\vV(t)|}\right]\frac12\ln\frac{1+\frac1c\abs{\vV(t)}}{1-\frac1c\abs{\vV(t)}} 
\right)} \right]
 +} O(\varkappa^3)
\end{alignat}
\end{widetext}
 The term in the first line at r.h.s.\refeq{eq:SELFforceEXPLICITexpand} is the contribution from the first line
at r.h.s.{\color{red}\refeq{pi2}}, the term in the second line at r.h.s.\refeq{eq:SELFforceEXPLICITexpand} is the contribution from the first line
at r.h.s.{\color{red}\refeq{pi1}}. 
 Since for straight-line motion $\vV(t)$ and $\aV(t)$ are parallel, and $\aV(t)=\dot\vV(t)$, one can carry out the time integration in 
the second line at r.h.s.\refeq{eq:SELFforceEXPLICITexpand} in terms of elementary functions of $\vV$, and a few algebraic manipulations 
then give \refeq{eq:SELFforceEXPLICITexpandMERGE}.
 To $O(\varkappa^2)$ this is the \emph{exact} expression for our BLTP radiation-reaction force 
in this problem where the particle starts from rest.
 It vanishes{\color{orange}.} 
{\color{orange}\sout{only as $|\vV|\downarrow 0$ and as $|\vV|\uparrow c$, and otherwise points against~$\vV$.}} 
{\color{magenta}A similar computation (to be published) reveals that the term $\propto\varkappa^3$ does not vanish.}
This demonstrates that the BLTP electrodynamical initial value problem accounts for the
radiation-reaction on the motion of a point charge along a uniform electric~field.

{\color{orange}\sout{In the vicinity of the initial time when $\frac{|\vV(t)|}{c}\ll 1$ we 
can expand and obtain, to leading order in ${|\vV|}/{c}$:}}\vspace{-5pt}
\begin{alignat}{1}
\label{eq:SELFforceEXPLICITexpandEXPAND}
{\color{orange}\cancel{\fV^{\mbox{\tiny{em}}}_1(t)}
\cancel{=}
\cancel{- \tfrac16 \tfrac{e^2 \varkappa^2}{c} \vV(t)}
\cancel{ \left[1+ O\left(\tfrac{|\vV(t)|^2}{c^2} \right) \right]}
\cancel{+} \cancel{O(\varkappa^3).}}
\end{alignat}
{\color{orange}\sout{Note that \refeq{eq:SELFforceEXPLICITexpandEXPAND} is a radiation-friction force
of the familiar ``Newtonian friction'' type (i.e. proportional to $-\vV$).}}

{\color{orange}\sout{In Fig.~1 we show $v$ in units of $c$ as a function of $t$ in units of $\mbare c / e^2\varkappa^2$
both for the test particle motion and for the BLTP motion with
radiation-friction force given by \refeq{eq:SELFforceEXPLICITexpandMERGE}.
 The applied electric field strength is $0.1$ in units of $e\varkappa^2$,  
which is a strong field for this problem.
The radiation-friction effect is clearly visible.}}

\begin{figure}[h]
{ \includegraphics[width = 6truecm,scale=2]{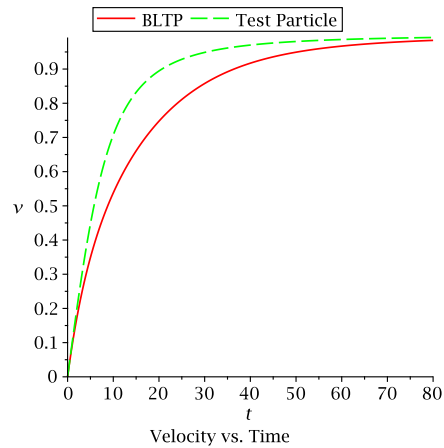} }
\caption{The velocity of a point charge, starting from rest in a strong, constant applied electrostatic field $\EV^{\mbox{\tiny{hom}}}$,
as per test particle theory (dashed curve), {\color{orange}\sout{and as per BLTP electrodynamics with $O(\varkappa^2)$ friction only (continuous curve).}}
\color{magenta}(Ignore the continuous curve.)}
\end{figure}

{\color{orange}\sout{ We note that for very weak applied field strength the radiation-reaction is captured by the ``Newtonian-friction'' 
approximation \refeq{eq:SELFforceEXPLICITexpandEXPAND}, and the point charge's velocity will saturate at 
$\vV_\infty= -\frac{6c}{e\varkappa^2}\EV^{\mbox{\tiny{hom}}}$; see Fig.~2 for an applied field strength
of $0.01$ in units of $e\varkappa^2$.}}

\begin{figure}[h]
  \includegraphics[width = 6truecm, scale=2]{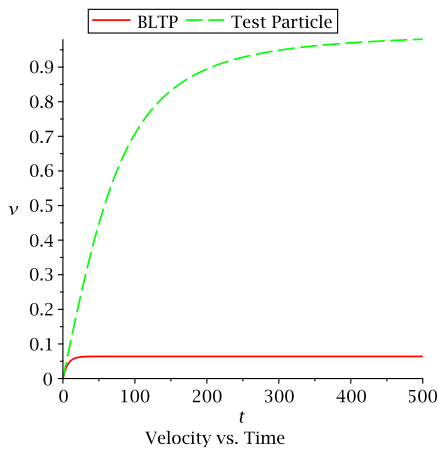} 
\caption{The velocity of a point charge, starting from rest in a very weak, constant applied electrostatic field $\EV^{\mbox{\tiny{hom}}}$,
as per test particle theory (dashed curve), {\color{orange}\sout{and as per BLTP electrodynamics with $O(\varkappa^2)$ friction only (continuous curve).}}
\color{magenta}(Ignore the continuous curve.)}
\end{figure}

{\color{orange}\sout{The sharp borderline between the ``weak field'' and ``strong field'' regimes determines a \emph{critical field strength} for this
problem. 
 For field strengths just slightly above the critical value $\approx 0.0519$ the velocity temporarily reaches a quasi-plateau, 
before it makes the final transition to approach the speed of light; see Fig.~3 for an applied field strength
of $0.052$ in units of $e\varkappa^2$.}}

\begin{figure}[h]
  \includegraphics[width = 6truecm,scale=2]{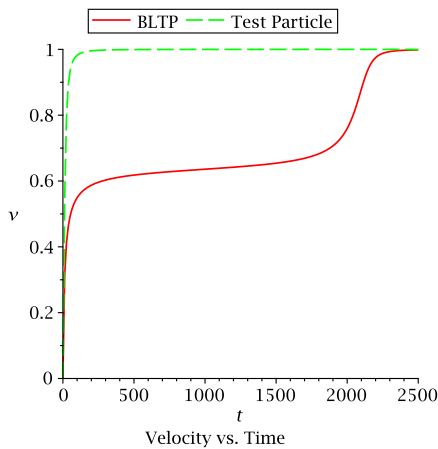} 
\caption{The velocity of a point charge, starting from rest in a
constant applied electrostatic field $\EV^{\mbox{\tiny{hom}}}$ {\color{orange}\sout{of slightly larger-than-critical field strength,}}
as per test particle theory (dashed curve), {\color{orange}\sout{and as per BLTP electrodynamics with $O(\varkappa^2)$ friction only (continuous curve).}}
\color{magenta}(Ignore the continuous curve.)}
\end{figure}

 Although the $\propto\varkappa^{\color{orange}3}$ term \refeq{eq:SELFforceEXPLICITexpandMERGE} is not an accurate formula for the radiation-reaction 
force in the physically interesting regime of very large $\varkappa$ values, it is entirely adequate for
demonstrating that the radiation-reaction does not vanish identically in the BLTP version of this standard textbook-type problem.
 Moreover, we have treated the dynamics properly as a physical initial value problem with the same data as in the test
particle formulation, unlike the treatment in \cite{Zayats} where
 ``the whole path traversed by the particle up to the present time contributes to [the self-force]'' (quoted from \cite{Zayats}).
\vspace{-5pt}

	\section{Summary and Outlook}\label{sec:CONCLUSIONS}\vspace{-5pt}
 In this paper we have shown that a well-defined electromagnetic force on a point charge source of the classical electromagnetic 
field can be extracted from momentum balance among charges and field, whenever the electromagnetic vacuum law which supplies the 
closure relation for the pre-metric Maxwell field equations leads to a finite field momentum vector which is differentiable with 
respect to time.

 For the BLTP law of the vacuum we even reported that we were able to prove,
in collaboration with S. Tahvildar-Zadeh, that the BLTP electromagnetic force as defined in \refeq{eq:EMf} 
furnishes a well-posed joint initial value problem for
fields and point particles which is of second-order in the particle positions; see \cite{KTZonBLTP} for the details.
 To the best of our knowledge BLTP electrodynamics is the first classical electrodynamical theory of point charges 
and their electromagnetic fields which has been shown to be dynamically well-posed, free of infinite ``self'' energies etc. 
and ill-defined Lorentz ``self'' forces, and free of the $\dddot\qV$-problem.
 Incidentally, neither Bopp, Land\'e--Thomas, nor Podolsky considered the definition of the force given in this paper,
but tried (in vain) to implement the ill-defined Lorentz force formula into their theory.

 We have illustrated the well-posed BLTP electrodynamical initial value problem by revisiting the standard textbook
problem of a point charge released from rest in a constant applied electrostatic field. 
 Our discussion confirms that the test-particle approximation is valid in the initial 
dynamical phase, with radiation-reaction corrections first in form of a linear {\color{orange}, \sout{``Newtonian friction''-type}}  term
(at least in the small $\varkappa$ regime), and eventually in a non-linear manner.
 
 A most interesting finding, valid for arbitrary $\varkappa$, 
is that in the initial phase of the dynamics the particle inertia is determined entirely by
its bare rest mass, not by the mass of the (electromagnetically) ``dressed'' particle. 
 The latter is generally thought to control the inertia in scattering scenarios.
 The predominance of scattering experiments, in particular in high energy physics, has led to the general belief that
only the ``dressed particle'' mass is observable in experiments, not the bare mass.
 Our findings by contrast suggest that the bare mass may be observable by cleverly setting up an initial value problem
in the laboratory.

 True, BLTP electrodynamics may not be the most realistic classical theory, but it surely is a ``proof of concept,''
signaling that analogous results should be feasible also for putatively more realistic models, in particular the BI electrodynamics.
 We remark that a well-defined joint initial value problem for the MBI fields and their point charge sources was formulated
with the help of a Hamilton--Jacobi-type theory in \cite{KieMBIinJSPa}, but it is not clear whether that theory is well-posed,
nor is it clear that its dynamics is independent of the invoked foliation of spacetime it needs for its formulation.
 Since the setup given in the present paper is truly Lorentz co-variant and foliation-independent, it
should shed light on the formulation given in \cite{KieMBIinJSPa} by comparing the two.
 Incidentally, neither Born \&\ Infeld, nor Schr\"odinger, nor Dirac, proposed the
electromagnetic force given in this paper but instead tried to implement the ill-defined Lorentz force 
formula into the Born--Infeld electrodynamics;~cf.~\cite{KieMBIinREGENSBURG}.

 Inside the family of well-posed classical models one may hope to find the classical limit of the elusive, 
mathematically well-defined and physically viable, special-relativistic quantum theory of electromagnetism. 

 Having obtained a rigorous control over  the classical electromagnetic radiation-reaction problem, an
important next goal in the realm of classical physics is to get a rigorous hand on the gravitational radiation-reaction problem.
 As a first step, armed with the insights gained from the special-relativistic theory of motion formulated in this paper
we have embarked on an assessment of the Einstein--Infeld--Hoffmann \cite{EIH} legacy; cf. \cite{KTZonEIHa}. 

\medskip
\section*{Acknowledgment}
The ideas and insights reported here evolved over many years, during which I have benefited 
especially from discussions with (in alphabetical order): Walter Appel, Holly Carley, Dirk Deckert, Detlef D\"urr, 
Vu Hoang, Markus Kunze, Volker Perlick, Maria Radosz, Jared Speck, Herbert Spohn, and Shadi Tahvildar-Zadeh
in particular, for which I am very grateful.
 After the first version of this paper was made public, the author received several comments which were implemented
in the revision. 
 Many thanks go: to Bob Wald for drawing attention to Ref. \cite{GHW}; to Mario Hubert for catching a sign error in the
Born--Infeld section; to Herb Johnson for prompting me to explain the MBLTP initial value problem more clearly; 
to the referee for the helpful comments which improved the presentation, and for prompting me to work out the 
example discussed in section VII.\vspace{-.5truecm}
\section*{References}

\bibliographystyle{apsrev}

\end{document}